%% file: sospaper.tex
\documentclass[pra,twocolumn]{revtex4-1}
\usepackage[utf8]{inputenc}
\usepackage{amsmath,amsfonts}
\usepackage{IEEEtrantools}
\usepackage{braket}

\usepackage[bookmarks=false]{hyperref}

\usepackage{graphicx}

\usepackage{float}

\newcommand{\vect}{\mathbf}

\newcommand{\mean}[1]{\mathopen{\langle}#1\mathclose{\rangle}}
\newcommand{\abs}[1]{\mathopen{\lvert}#1\mathclose{\rvert}}
\newcommand{\norm}[1]{\mathopen{\lVert}#1\mathclose{\rVert}}
\newcommand{\Norm}[1]{\left\lVert#1\right\rVert}

\newcommand{\cointerval}[1]{\mathopen{[}#1\mathclose{[}}

\newcommand{\ocinterval}[1]{\mathopen{]}#1\mathclose{]}}
\newcommand{\ccinterval}[1]{\mathopen{[}#1\mathclose{]}}

\DeclareMathOperator{\tr}{Tr}
\DeclareMathOperator{\vspan}{span}
\DeclareMathOperator{\End}{End}
\newcommand{\bigo}{O}

\newcommand{\id}{{\mathbb I}}
\newcommand{\iop}{\mathcal I}
\newcommand{\ibar}{\overline{\iop}}

\newcommand{\imax}{I^{\mathrm{max}}}

\newcommand{\monomialS}{\mathcal S}

\newcommand{\eqp}{\,}

\newcommand{\rma}{\mathrm{a}}
\newcommand{\rmA}{\mathrm{A}}
\newcommand{\rmB}{\mathrm{B}}

\newcommand{\junk}{\mathrm{junk}}

\newcommand{\apdxref}[1]{Appendix~\ref{apdx:#1}}

\begin{document}
\title{Sum-of-squares decompositions for a family of CHSH-like inequalities\\ and their application to self-testing}

\author{C\'edric Bamps}
\author{Stefano Pironio}
\affiliation{Laboratoire d'Information Quantique, Universit\'e libre de Bruxelles (ULB), 1050 Brussels, Belgium}

\begin{abstract}
We introduce two families of sum-of-squares (SOS) decompositions for the Bell operators associated with the tilted CHSH expressions introduced in \emph{Phys.\ Rev.\ Lett.\ 108, 100402 (2012)}.
These SOS decompositions provide tight upper bounds on the maximal quantum value of these Bell expressions.
Moreover, they establish algebraic relations that are necessarily satisfied by any quantum state and observables yielding the optimal quantum value.
These algebraic relations are then used to show that the tilted CHSH expressions provide robust self-tests for any partially entangled two-qubit state.
This application to self-testing follows closely the approach of \emph{Phys.\ Rev.\ A 87, 050102 (2013)}, where we identify and correct two non-trivial flaws.
\end{abstract}

\maketitle

\section{Introduction}
We consider a bipartite Bell scenario \cite{rmp} with two binary observables $A_0$, $A_1$ for Alice and two binary observables $B_0$, $B_1$ for Bob, such that $A_x^2 = \id$, $B_y^2 = \id$, and $[A_x,B_y]=0$ for all $x,y=0,1$ \footnote{We assume implicitly that Alice's observables are of the form $A_x\otimes \id$ and those of Bob of the form $\id\otimes B_y$.}.
The following family of tilted CHSH operators for this scenario was introduced in \cite{acin_randomness_2012}:
\begin{equation}\label{ia}
\iop_\alpha = \alpha A_0 + A_0B_0 + A_0B_1 + A_1B_0 - A_1B_1 \eqp,
\end{equation}
where $\alpha \in \cointerval{0,2}$ is a parameter and $\alpha=0$ corresponds to the CHSH operator.
One of our objectives in this article is to derive an upper bound $\eta_\alpha$ on the maximal quantum expectation value of $\iop_\alpha$, that is to show that $\mean{\iop_\alpha} \le \eta_\alpha$ for all possible quantum states and measurement operators $A_x$ and $B_y$.
This is equivalent to showing that the operator $\ibar_\alpha = \eta_\alpha \id - \iop_\alpha$ is positive semidefinite, i.e.\ $\ibar_\alpha \succeq 0$, for any measurement operators $A_x$ and $B_y$.
This in turn can be proven by providing a set of operators $\{P_i\}$ which are polynomial functions of $A_x$ and $B_y$ such that
\begin{equation}
\ibar_\alpha = \sum_i P_i^\dag P_i^{\vphantom\dag}
\label{eq:sos}
\end{equation}
holds for any set of measurement operators satisfying the algebraic properties $A_x^2 = \id$, $B_y^2 = \id$, and ${[A_x,B_y]=0}$.
Indeed, if $\ibar_\alpha$ is of this form, it is obviously positive semidefinite.
A decomposition of the form \eqref{eq:sos} is called a \emph{sum of squares} and can be defined in full generality for other Bell operators than the ones considered here.

Though it may be difficult to find a SOS decomposition for an arbitrary Bell operator, especially one yielding a tight bound $\eta_\alpha$ on the maximal quantum violation, once one has been found, verifying that \eqref{eq:sos} holds, and thus that $\mean{\iop_\alpha} \le \eta_\alpha$, usually involves only a few simple computations.
That is, a SOS provides a simple certificate that $\mean{\iop_\alpha} \le \eta_\alpha$.
Furthermore, the search for optimal SOS can be cast as a series of semidefinite programs (SDP) that turns out to be simply the dual formulation \cite{doherty} of the SDP hierarchy introduced in \cite{npa, navascues_convergent_2008}.
Finally, as shown in \citep{yang_robust_2013}, an optimal SOS, i.e.\ one for which $\mean{\iop_\alpha} \le \eta_\alpha$ is a tight bound, provides useful information about the optimal quantum strategy and can find an application in robust self-testing.

Self-testing is the process through which one can guarantee that two devices satisfy certain properties, e.g.\ that they implement measurements on a quantum state which is close, up to a local isometry, to a given reference state, only by observing the correlations in a Bell experiment \cite{mayers}.
The possibility of self-testing means that the interaction on a classical level with quantum devices can be sufficient to assure users that they indeed hold devices that conform to an ideal specification: as long as the self-testing criteria are satisfied, it is guaranteed that the devices have not malfunctioned or been tampered with.
Effectively, this allows the users to treat their devices as black boxes.
This is the core idea of device-independent quantum information processing, where self-testing has been used as a primitive to establish schemes for verified quantum computing \cite{ruv} and cryptographic tasks such as randomness expansion \cite{mayers,miller_robust_2014}.

Despite their interest, very few examples of explicit SOS decompositions for Bell operators have been given in the literature.
One example is a SOS for the family of ``guess your neighbour's input" inequalities introduced in \cite{GYNI}.
A second example is a SOS for the CHSH operator appearing in \cite{ruv}, which we recover in this article.
Finally, SOS decompositions for the entire family of tilted CHSH operators \eqref{ia} were proposed in \cite{yang_robust_2013}.
However, as we point out here, these SOS only hold in the range $\alpha \in [0.156,1.955]$; outside that range they are not valid SOS due to a sign error that cannot be simply fixed.
Furthermore, even within the validity range, these SOS decompositions are not sufficient for the self-testing application proposed in \cite{yang_robust_2013} (see \apdxref{shortcomings}).

In this article, we introduce two different simple SOS decompositions for the tilted CHSH operators \eqref{ia} that are valid for the entire range $\alpha \in \cointerval{0,2}$, including the CHSH case $\alpha=0$.
Our systematic approach to this problem is of independent interest as it can probably be adapted to other Bell operators.
We then show, following \cite{mckague_robust_2012} and \cite{yang_robust_2013}, how to apply these SOS decompositions to robust self-testing of any partially entangled state.
Moreover, we take the self-testing analysis further by showing that the isometry providing a robust self-test for the state, also provides a self-test for the action of the measurement operators.
Incidentally, we point out and fix a small mistake in \citep{mckague_robust_2012} concerning the regularization procedure used to define the local isometry used for self-testing.

This article is organized as follows.
We first present our SOS decompositions and detail the approach we used to obtain them.
We then sketch their application to self-testing, which is comprehensively discussed in \apdxref{selftest}.
The flaws that we identified in \cite{yang_robust_2013} and \citep{mckague_robust_2012} and how our work resolves them is presented in \apdxref{shortcomings}.

\section{SOS decompositions for tilted CHSH inequalities}
We first start by reviewing some of the properties of the tilted CHSH expressions $\iop_\alpha$.
We then introduce the concept of SOS decompositions and derive such decompositions for the tilted CHSH operators.

\subsection{Optimal quantum strategies for $\iop_\alpha$}
\label{sec:reference}
The optimal quantum value for the Bell expressions \eqref{ia} was computed in \cite{acin_randomness_2012} by optimizing explicitly over all quantum states and measurements.
The optimal value is $\imax_\alpha = \sqrt{8+2\alpha^2}$, to be compared to the classical bound $2 + \alpha$.
An interesting property of this class of Bell operators is that for all values of $\alpha \in \cointerval{0,2}$, the maximal quantum value can be achieved by a partially entangled qubit pair
\begin{equation}
\label{eq:refstate}
\ket{\psi} = \cos\theta \ket{00} + \sin\theta \ket{11}
\end{equation}
with $\alpha \equiv \alpha(\theta) = 2/\sqrt{1+2\tan^2 2\theta}$, effectively covering the full range of partial entanglement in two-qubit states, $\theta \in \ocinterval{0,\pi/4}$.
The operators used to achieve this maximal quantum violation are
\begin{IEEEeqnarray}{rL'rL"}
\IEEEyesnumber\phantomsection\label{eq:maxops}
\IEEEyessubnumber
   A_0 &= \sigma_z
&  A_1 &= \sigma_x
\\
\IEEEyessubnumber
   B_0 &= \cos\mu \>\sigma_z + \sin\mu \>\sigma_x
&  B_1 &= \cos\mu \>\sigma_z - \sin\mu \>\sigma_x
\end{IEEEeqnarray}
where $\tan\mu = \sin 2\theta$ and $\sigma_{x,z}$ are the $x$, $z$ Pauli operators.

\subsection{SOS decompositions for Bell operators}
Our goal is to find a decomposition as in \eqref{eq:sos} in terms of a set of polynomials $\{P_i\}$ when the constant term in $\ibar_\alpha$ is the maximal quantum value of the tilted CHSH operator, $\eta_\alpha = \imax_\alpha$.
For simplicity, we restrict our search space to the span of a small canonical basis of nine monomials of degree two,
\begin{equation}
\monomialS_{1+AB} = \{ \id, A_0, A_1 \} \otimes \{ \id, B_0, B_1 \} \eqp.
\end{equation}
To simplify the search, we will later pick a different basis of this vector space of polynomials, $\{R_i\}_i$.
Thus, the operators $P_i$ are decomposed as $P_i = q_i^\mu R_\mu$ (using implicit summation on repeated indices).
The expression of $\ibar_\alpha$ as an SOS \eqref{eq:sos} then becomes
\begin{equation}
\ibar_\alpha = R^\dag_\mu {q^\mu_i}^* q^\nu_i R_\nu^{\vphantom\dag} \equiv R^\dag_\mu M^{\mu\nu} R_\nu^{\vphantom\dag} \eqp.
\label{eq:psdform}
\end{equation}
If $q^\nu_i$ is now seen as the $i$th component of a vector $\vect q^\nu$, the hermitian matrix $M$ is then the Gram matrix of this set of nine vectors and is therefore positive semidefinite.
The converse is also true: if $M$ is positive semidefinite, there exists a (non-unique) set of vectors $\{\vect q^\nu\}$ such that the components $M^{\mu\nu}$ are the hermitian products $(\vect q^\mu, \vect q^\nu)={q^\mu_i}^* q^\nu_i$,
and therefore any operator of the form of the RHS of \eqref{eq:psdform} with $M \succeq 0$ is a sum of squares.
A set of such vectors $\{\vect q^\nu\}$ is given by the columns of any matrix square root of $M$, such as its Cholesky decomposition if it is nonsingular.

We are now looking for a positive semidefinite matrix $M$ such that \eqref{eq:psdform} holds.
This equation imposes linear constraints on $M$.
Indeed, we can decompose both sides of the equality $\ibar_\alpha = M^{\mu\nu} R^\dag_\mu R_\nu^{\vphantom\dag}$ in a basis of the quadratic products of all elements in $\monomialS_{1+AB}$, which is of size $25$ rather than $81$ due to the algebraic relations satisfied by the measurement operators.
A canonical basis for these products is
\begin{multline}
\monomialS_{1+AB}^2 = \{ \id, A_0, A_1, A_0A_1, A_1A_0 \} \\ {} \otimes \{ \id, B_0, B_1, B_0B_1, B_1B_0 \} \eqp.
\end{multline}
Writing $R^\dag_\mu R_\nu^{\vphantom\dag} = F_{\mu\nu}^i E_i^{\vphantom i}$ where $E_i$ runs through $S_{1+AB}^2$ and each $F^i$ is a matrix of complex coefficients, and likewise $\ibar_\alpha = s^i E_i$, the SOS condition \eqref{eq:psdform} reduces to
\begin{equation}
s^i = \tr(M^\dag F^i) \qquad (i=1,\dotsc,25) \eqp.
\label{eq:lincons}
\end{equation}
We are thus left with a set of 25 linear equality constraints on $M$ as well as the positive semidefiniteness constraint $M \succeq 0$.
This is a semidefinite programming feasibility problem, which can be approached with numerical tools such as SeDuMi \cite{sedumi}.

However, while we could attempt to recover the exact analytical expression that the numerical solution approximates, this will not be a good approach for a continuous class of Bell operators.
We will thus tackle this problem analytically, and this requires that we simplify the problem as much as possible.
One first simplification comes from our knowledge of a state and measurements which achieve the quantum bound.
One such system is specified in Eqs.~\eqref{eq:refstate} and \eqref{eq:maxops}, consisting in the partially entangled qubit pair $\ket{\psi} = \cos\theta \ket{00} + \sin\theta \ket{11}$ with spin measurements along given axes.
Because this strategy achieves the quantum bound $\imax_\alpha$, the expectation value of $\ibar_\alpha$ vanishes.
As a consequence, any SOS decomposition for $\ibar_\alpha$ of the form \eqref{eq:sos} must have each of its terms vanish in expectation as well.
Hence, a valid SOS decomposition for $\ibar_\alpha$ must be made up of terms for which $P_i \ket{\psi} = 0$ in this maximally violating quantum system.
This last equation defines four constraints that all $P_i$ must obey, one per basis vector of the Hilbert space.
Indeed, writing the most general $P$ in our search space as $\vect r \cdot \vect V$ where
\begin{equation}
\vect V = (\id, A_0, A_1, B_0, B_1, A_0B_0, A_0B_1, A_1B_0, A_1B_1)
\label{eq:vops}
\end{equation}
and demanding that the four components of $P \ket\psi = P(\cos\theta \ket{00} + \sin\theta \ket{11})$ (with the observables specified in \eqref{eq:maxops}) vanish, we find four independent linear constraints on the vector $\vect r$.
We thus find that the space of candidates $P_i$ is spanned by the following five operators:
\begin{subequations}
\label{eq:soscandidates}
\begin{gather}
Z_\rmA - Z_\rmB                            \eqp, \\
\id - Z_\rmA Z_\rmB                        \eqp, \\
c X_\rmA - s Z_\rmA X_\rmB - X_\rmA Z_\rmB \eqp, \\
c X_\rmB - s X_\rmA Z_\rmB - Z_\rmA X_\rmB \eqp, \\
s X_\rmA X_\rmB - Z_\rmA Z_\rmB + c Z_\rmA \eqp,
\end{gather}
\end{subequations}
where $c = \cos 2\theta$, $s = \sin 2\theta$, and the $Z$ and $X$ operators are defined as
\begin{IEEEeqnarray}{rL"rL}
\IEEEyesnumber\phantomsection\label{eq:xzgeneric}
\IEEEyessubnumber
   Z_\rmA &= A_0
&  X_\rmA &= A_1
\eqp,\\
\IEEEyessubnumber
   Z_\rmB &= \frac{B_0 + B_1}{2\cos\mu}
&  X_\rmB &= \frac{B_0 - B_1}{2\sin\mu}
\eqp.
\end{IEEEeqnarray}
It is easily verified that these five operators indeed vanish in the explicit two-qubit system we considered, as $Z$ and $X$ are then the Pauli operators.

This first step lets us write the $P_i$ polynomials as linear combinations of five operators instead of nine, hence the $M$ matrix in our SDP feasibility problem is now $5\times5$.

One further simplification comes from exploiting the symmetry of the tilted CHSH operator to impose a similar symmetry on its SOS decompositions.
The effect of symmetries on SOS decompositions has been studied in the case of polynomials of commutative variables in \cite{gatermann_symmetry_2004}, and the following takes inspiration from those results.
We observe that changing the sign of $A_1 \to -A_1$ and swapping $B_0 \leftrightarrow B_1$ leaves $\iop_\alpha$ (and therefore $\ibar_\alpha$) invariant.
This transformation induces a representation of the cyclic group $C_2$ on the vector space of operators $\vspan(\monomialS_{1+AB})$.
Let $\sigma \in \End(\vspan(\monomialS_{1+AB}))$ be the endomorphism representing the transformation.
Due to the invariance of the tilted CHSH operator, the five-dimensional subspace spanned by operators \eqref{eq:soscandidates} is itself invariant under $\sigma$ by definition.
Therefore, we separate this subspace as a direct sum of irreducible representation subspaces, all of dimension one according to the representation theory of cyclic groups.
We then choose our basis operators $\{R_i\}_{i=1}^5$ to be the basis elements associated with such a decomposition.
This means that $\sigma$ acts on this basis as $\sigma(R_i) = \pm R_i$, depending on $i$.

We now apply this symmetry transformation to both sides of \eqref{eq:sos}:
\begin{equation}
\ibar_\alpha = \sigma(\ibar_\alpha) = \sigma(R_\mu)^\dag M^{\mu\nu} \sigma(R_\nu) = R^\dag_\mu M'^{\mu\nu} R_\nu^{\vphantom\dag} \eqp,
\end{equation}
where $M'^{\mu\nu} = \pm M^{\mu\nu}$ depending on the sign brought by the transformation of the basis operators.
This means that $M'$ represents a new SOS decomposition for $\ibar_\alpha$.
Moreover, because convex combinations of SOS decompositions are also valid SOS decompositions of the same operator, a third SOS decomposition is found: $M^{(\mathrm S)} = (M + M')/2$.
This last SOS has the property of being invariant under the symmetry transformation.

While asymmetric SOS may still exist, this result is useful because it lets us focus on symmetric decompositions, which have a smaller number of degrees of freedom.
Indeed, because elements in $M'$ and $M$ only differ by their sign, the symmetrization of $M$ to $M^{(\mathrm S)}$ will take to zero the elements that change sign.
The elements in question correspond to indices $(\mu,\nu)$ such that $R_\mu$ and $R_\nu$ span representation subspaces of different irreducible representations of $C_2$.
As a result, the symmetrized SOS matrix $M^{(\mathrm S)}$ is block-diagonal, with one block associated to each of the two irreducible representations.

Considering the discussion thus far, we now choose a basis for the subspace containing the SOS polynomials.
The basis we choose is $\{ R_i \equiv \vect r_i \cdot \vect V \}$ where the $\vect r_i$ vectors are defined as such (we label the columns with the operators defining $\vect V$ for convenience):
\begin{widetext}
\begin{equation}
\newcommand{\sqs}{\sqrt{1+s^2}}%
\newcommand{\df}{\dfrac}%
\renewcommand{\arraystretch}{2}%
\begin{array}{c@{}ccccccccc@{}c}
                  & \id           & A_0           & A_1            & B_0 & B_1 & A_0B_0 & A_0B_1 & A_1B_0 & A_1B_1 & \\
\vect r_1 = \Big( & 0             & \df{-2}{\sqs} & 0              & 1   & 1   & 0      & 0      & 0      & 0      & \Big), \\
\vect r_2 = \Big( & \df{-2}{\sqs} & 0             & 0              & 0   & 0   & 1      & 1      & 0      & 0      & \Big), \\
\vect r_3 = \Big( & \df{-2}{\sqs} & 0             & 0              & c   & c   & 0      & 0      & 1      & -1     & \Big), \\
\vect r_4 = \Big( & 0             & 0             & \df{ -2}{\sqs} & 1   & -1  & 0      & 0      & c      & c      & \Big), \\
\vect r_5 = \Big( & 0             & 0             & \df{-2c}{\sqs} & 0   & 0   & 1      & -1     & 1      & 1      & \Big).
\end{array}
\label{eq:newr}
\end{equation}
\end{widetext}
It is easily checked that these basis operators separate the space in two isotypical subspaces, i.e., subspaces that fall under the same irreducible representation of the cyclic group: $R_{1,2,3}$ are invariant under the symmetry transformation of $\iop_\alpha$, while $R_{4,5}$ change sign.
The block structure of symmetric SOS matrices is therefore $3 \oplus 2$, where the first block corresponds to the trivial representation and the second to the parity representation where the group generator is represented by $-1$: we have $\sigma(R_i) = R_i$ for $i = 1,2,3$ and $\sigma(R_i) = -R_i$ for $i = 4,5$.

We now examine the problem of finding SOS decompositions with $\alpha = 0$, where $\iop_0$ is the CHSH operator, and with $\alpha = \alpha(\pi/8)$.

\subsubsection{The CHSH case: $\alpha = 0$}
\begin{figure}[H]
\centering%
%\vspace{5mm}%
\includegraphics[resolution=300]{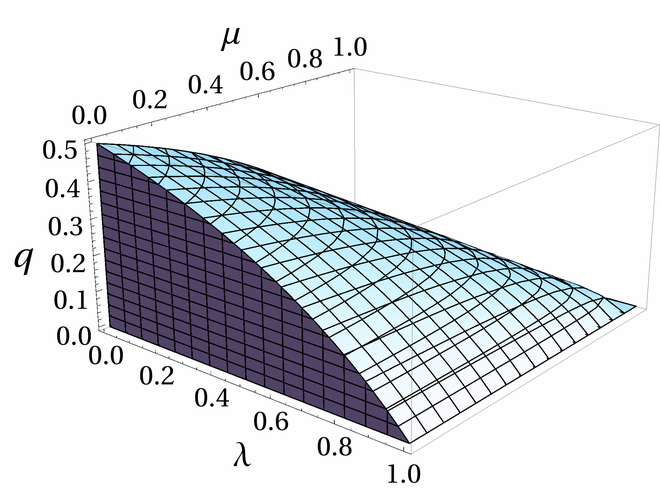}
\caption{\label{fig:chshvol}%
Solution set for symmetric SOS decompositions for the CHSH operator in $\mathcal S_{1+AB}$.
}
\end{figure}

We first look at the determination of SOS decompositions in the simplest case, where $\alpha = 0$.
Without yet taking symmetry into account, the linear constraints imposed on $M$ by Eq.~\eqref{eq:lincons} imply
(omitting the lower triangular part given by hermiticity)
\begin{equation}
M = \frac{1}{2\sqrt{2}}
\begin{pmatrix}
\lambda & 0      & 0 & \delta & 0         \\
        & \gamma & q & 0      & -\delta/2 \\
        &        & t & 0      & -\delta/2 \\
        &        &   & \mu    & 0         \\
        &        &   &        & q
\end{pmatrix}
\label{eq:chshsosmatrix}
\end{equation}
with $q = 1 - \lambda - \gamma$ and $t = \lambda + \gamma - \mu$.
Thus, $M$ depends on four parameters, three of which are real due to hermiticity.
The fourth parameter, $\delta$, can be taken to $0$ by the symmetry argument given above.
We see that, in this specific case, the $1 \oplus 2 \oplus 1 \oplus 1$ block structure of a symmetric $M$ goes further than what is imposed by the symmetry of $\iop_\alpha$.
This is not surprising, as the CHSH operator, lacking a marginal term, is more symmetric than the general $I_\alpha$ operator.
This makes it straightforward to impose positive semidefiniteness: a necessary and sufficient condition is the positive semidefiniteness of the blocks.
We find the conditions
\begin{gather}
\lambda, \mu, q, t \ge 0 \eqp,
\\
\label{eq:chshquad}
\lambda t - q^2 \ge 0 \eqp.
\end{gather}
We therefore end up with a simple description of the full solution set of symmetric SOS decompositions of $\ibar_0$ in $\monomialS_{1+AB}$, up to the computation of a square root for $M$ to get the actual polynomials $P_i$ in \eqref{eq:sos}.
We represent this solution set in terms of $\lambda$, $\mu$ and $q$ in Figure~\ref{fig:chshvol}.

In this solution set, the extremal points are the most interesting as they generate the entire set by convex combinations.
Here, the extremal points are the five vertices of the set and the smooth quadratic surface resulting from inequality \eqref{eq:chshquad}.
Because of their situation at the intersection of inequality constraints, the $M$ matrices at the five vertices are of low rank, which eases the determination of a square root.
Those distinguished extremal points are listed in Table~\ref{tab:a:chshvertexsos}.
We note that the first four vertex SOS in the table are equivalent up to multiplication on the left by a dichotomic operator on one or two of the operators that are squared in the sum.
For instance, the identity $R_2=A_0R_1$ implies that $R_2^2=R_1^2$, which means that the SOS $\mathrm{C}_1$ and $\mathrm{C}_2$ are equivalent.

We reproduce here two SOS decompositions of $\ibar_0$ resulting from those extremal points:
\begin{gather}
\ibar_0 = \frac{1}{4\sqrt{2}} \left[ \ibar_0^2 + 2 ( Z_\rmA X_\rmB + X_\rmA Z_\rmB )^2 \right]
\label{eq:chshsos1}
\\
\ibar_0 = \frac{1}{\sqrt{2}} \left[ (-Z_\rmA + Z_\rmB)^2 + (-X_\rmA + X_\rmB)^2 \right]
\label{eq:chshsos2}
\end{gather}
where $Z_\rmA = A_0$, $X_\rmA = A_1$, $Z_\rmB = (B_0+B_1)/\sqrt{2}$, and $X_\rmB = (B_0-B_1)/\sqrt{2}$.
We note that the decomposition \eqref{eq:chshsos2}, which we denote as $\mathrm{C}_4$ in Table~\ref{tab:a:chshvertexsos}, also appears in \cite{ruv}.

Two additional SOS decompositions for CHSH are also given in \apdxref{extrachsh}.

\subsubsection{The $\theta = \pi/8$ case}
Next, we choose the (arbitrary) value of $\theta = \pi/8$.
The equality constraints \eqref{eq:lincons} give to the SOS matrix the following form:
\begin{widetext}
\begin{equation}
M =
\frac{\sqrt{3}}{8\sqrt{2}}
\begin{pmatrix}
\beta & \frac{-1}{\sqrt2}               & \frac{3}{\sqrt2}\gamma & 3\delta       & 0                                \\
      & \frac53 - \beta + \frac12\gamma & \frac13 - 2\gamma      & -\sqrt2\delta & -\delta                          \\
      &                                 & 2 + 3\gamma - \lambda  & \sqrt2\delta  & -2\delta                         \\
      &                                 &                        & \lambda       & \frac{-\sqrt2}{3} - \sqrt2\gamma \\
      &                                 &                        &               & \frac23 - \gamma                 %
\end{pmatrix} \eqp,
\label{eq:pi8sosmatrix}
\end{equation}
\end{widetext}
where, again, $\delta$ can be set to zero by symmetry to give $M$ a block-diagonal structure.

\begin{figure}[t]
\includegraphics[resolution=300]{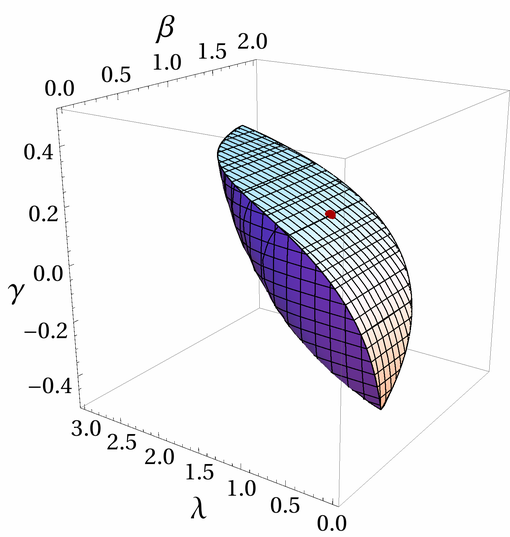}
\caption{\label{fig:pi8vol}%
Solution set for symmetric SOS decompositions for $\iop_{\alpha(\pi/8)}$ in $\mathcal S_{1+AB}$.
This set is delimited by a quadratic surface (invariant along $\beta$) and a cubic surface.
Yang and Navascu\'es' SOS \cite{yang_robust_2013} is represented as a red dot.
}
\end{figure}

Contrary to the CHSH matrix, the block structure in this case is coarser with fewer, larger blocks, which results in more nonlinear positive semidefiniteness constraints.
The positive semidefiniteness of $M$ is checked by using a generalized form of Sylvester's criterion, namely that a matrix is positive semidefinite if and only if all of its principal minors (submatrices obtained by jointly eliminating lines and columns of the same indices), including the matrix itself, have a nonnegative determinant \cite{bhatia_positive_2009}.
The solution set is visualized as a three-dimensional convex region in Figure~\ref{fig:pi8vol}, delimited by a quadratic surface and a cubic surface.

We identify on Figure~\ref{fig:pi8vol} two points that stand out, namely the two cusps at the intersection of the two surfaces delimiting the region.
Their coordinates are found to be $(\beta,\gamma,\lambda) = (1/2,-1/3,0)$ and $(3/2,1/3,8/3)$.
Both blocks of $M$ have rank one at these points, which eases the computation of square roots $N$ that describe the SOS polynomials.
We find respectively
\begin{gather}
N_1 = \sqrt{\frac{\sqrt3}{8\sqrt2}}
\begin{pmatrix}
1/\sqrt2 & -1 & -1 & 0 & 0  \\
0        & 0  & 0  & 0 & -1
\end{pmatrix}
\eqp, \\
N_2 = \sqrt{\frac{\sqrt3}{8\sqrt2}}
\begin{pmatrix}
-\sqrt{\frac32} & \frac{1}{\sqrt3} & -\frac{1}{\sqrt3} & 0                & 0                \\
0               & 0                & 0                 & -2\sqrt{\frac23} & \frac{1}{\sqrt3}
\end{pmatrix}
\eqp.
\end{gather}
The rows in these matrices give us the coefficient vectors $\vect q_i$ for two explicit SOS decompositions of $\ibar_{\alpha(\pi/8)}$.

\subsubsection{General tilted CHSH}
The general approach is the same as above: first, the linear constraints \eqref{eq:lincons} are imposed, and imposing symmetry leaves us with three real degrees of freedom.
The positive semidefiniteness of the blocks of $M$ is then enforced by applying the generalized Sylvester's criterion \cite{bhatia_positive_2009}.
This amounts to one cubic, four quadratic and five linear inequalities.
For a given value of $\alpha$, the solution set is readily visualized as a convex set in $\mathbb R^3$.

In the previous section, we examined the case of $\alpha = \alpha(\pi/8)$ and identified two extremal points in the solution set where each block of the SOS matrix $M$ is of rank one.
When we carry out the explicit expansion of the sums of squares, these decompositions hint at an easy generalization to all values of $\alpha$ for both points.
This leads us to the two SOS decompositions that follow.
Let us first define the following permutations of the CHSH operator:
\begin{align}
\mathcal S'   &= A_0 (B_0-B_1) + A_1 (B_0+B_1) \eqp, \\
\mathcal S''  &= A_0 (B_0+B_1) - A_1 (B_0-B_1) \eqp, \\
\mathcal S''' &= A_0 (B_0-B_1) - A_1 (B_0+B_1) \eqp.
\end{align}
The two decompositions we find are the following:
\begin{equation}
\ibar_\alpha = \frac{1}{2\imax_\alpha} \left[ \ibar_\alpha^2 + \bigl(\alpha A_1 - \mathcal S'\bigr)^2 \right]
\label{eq:sos1}
\end{equation}
and
\begin{multline}
\ibar_\alpha = \frac{1}{2\imax_\alpha}
 \Bigl[
      \bigl( 2 A_0 - \imax_\alpha \frac{B_0+B_1}{2} + \frac{\alpha}{2} \mathcal S''  \bigr)^2
  \\ {}+\bigl( 2 A_1 - \imax_\alpha \frac{B_0-B_1}{2} + \frac{\alpha}{2} \mathcal S''' \bigr)^2
 \Bigr] \eqp.
\label{eq:sos2}
\end{multline}
The two SOS decompositions given in the CHSH case \eqref{eq:chshsos1}, \eqref{eq:chshsos2} are limiting cases of \eqref{eq:sos1} and \eqref{eq:sos2}.
Independently of the way we arrived at \eqref{eq:sos1} and \eqref{eq:sos2}, it is readily verified that they correspond to valid SOS.

\section{Application to self-testing}
We conclude by discussing the application of the above results to self-testing.
We know that measuring the state $\ket{\psi}=\cos\theta \ket{00} + \sin\theta\ket{11}$ with the observables \eqref{eq:maxops} leads to the maximal value $\imax_\alpha$ of the Bell expression $\iop_\alpha$.
We say that this Bell expression provides a robust self-test for this particular reference state and reference observables if the converse also holds, in a noise-tolerant way --- that is, if an expected value of $\iop_\alpha$ close to the maximum $\imax_\alpha$ necessarily corresponds to measurements involving a state and observables that are close to the reference, up to a local isometry.
More precisely, we show, following the framework of \cite{mckague_robust_2012} and \cite{yang_robust_2013}, the following.
Let $\langle \tilde \iop_\alpha\rangle$ be the expectation value of the Bell expression $\iop_\alpha$ obtained by measuring a physical state $\ket{\tilde\psi}$ with physical observables $\tilde A_x$ and $\tilde B_y$.
Let $\ket{\psi}$ and $A_x$ and $B_y$ be the reference state and reference observables corresponding to the optimal quantum strategies for $\iop_\alpha$ defined in Sec.~\ref{sec:reference}.
Then if $\langle \tilde\iop_\alpha\rangle\geq \imax_\alpha-\epsilon$, there exists a local isometry $\Phi = \Phi_\rmA \otimes \Phi_\rmB$ and a state $\ket{\junk}$ such that
\begin{equation}
\norm{ \Phi(\tilde A_x \otimes \tilde B_y \ket{\tilde\psi}) - \ket{\junk} \otimes (A_x \otimes B_y) \ket{\psi} } \leq \epsilon'
\label{eq:selftest}
\end{equation}
for $x, y \in \{-1,0,1\}$, where the subscript $-1$ refers to the identity operator and where $\epsilon'=\bigo(\sqrt{\epsilon})$.
The precise relation between $\epsilon$ and $\epsilon'$ is given in \apdxref{selftest}.

The case of the CHSH inequality ($\alpha=0$) was already considered in \cite{mckague_robust_2012}, where Eq.~\eqref{eq:selftest} was obtained but with a weaker $\epsilon'=\bigo(\epsilon^{1/4})$ robustness.
Equation \eqref{eq:selftest} in the case $x,y=-1$ for arbitrary values of $\alpha\in \cointerval{0,2}$, corresponding to a self-test of any partially entangled state (but not of the associated observables) was considered in \cite{yang_robust_2013}, where the authors showed that the use of SOS decompositions could in principle lead to an optimal $\epsilon'=\bigo(\sqrt\epsilon)$ robustness.
This conclusion, however, does not follow from the analysis presented in \cite{yang_robust_2013} due to different shortcomings in the derivation of intermediate results (see \apdxref{shortcomings} for details).
Here we resolve these shortcomings, thanks in particular to the two SOS decompositions \eqref{eq:sos1} and \eqref{eq:sos2}, and establish the robust self-testing conditions \eqref{eq:selftest}, thus further extending the analysis in \cite{yang_robust_2013} to the self-testing of the observables \eqref{eq:maxops} in addition to the state \eqref{eq:refstate}.

\begin{figure}[t]
\includegraphics{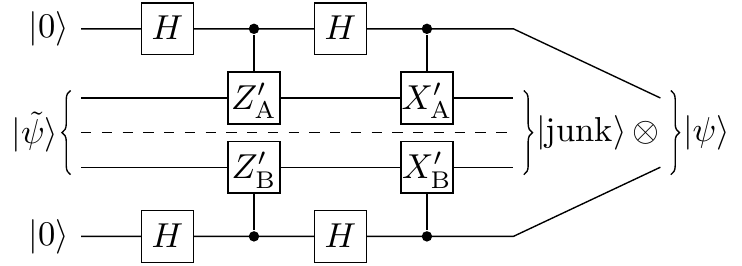}
\caption{\label{fig:circuit}%
The local isometry used to locate the partially entangled qubit pair in the physical system.
The dashed line shows the physical separation between Alice (top) and Bob (bottom).
}
\end{figure}

Following closely \cite{yang_robust_2013}, we now sketch the proof of the robust self-testing result \eqref{eq:selftest} and refer to \apdxref{selftest} for details.
The proof proceeds in three steps.
First, we define as in \cite{mckague_robust_2012} the isometry appearing in \eqref{eq:selftest} as the successive action of a set of gates, represented as a circuit in Figure~\ref{fig:circuit}, acting on the initial state $\ket{00}\otimes\ket{\tilde \psi}$.
The gates are defined in terms of the measurement operators $\tilde A_x$ and $\tilde B_y$ as an inversion of \eqref{eq:maxops} in an attempt to recover the behaviour of the Pauli operators.
Specifically, we define as in \eqref{eq:xzgeneric}
\begin{IEEEeqnarray}{rL"rL}
\IEEEyesnumber\phantomsection\label{eq:xz}
\IEEEyessubnumber
   \tilde Z_\rmA &= \tilde A_0
&  \tilde X_\rmA &= \tilde  A_1
\eqp,\\
\IEEEyessubnumber
   \tilde Z_\rmB &= \frac{\tilde B_0 + \tilde B_1}{2\cos\mu}
&  \tilde X_\rmB &= \frac{\tilde B_0 - \tilde B_1}{2\sin\mu}
\eqp.
\end{IEEEeqnarray}
Because we need unitary gates in the isometry, a regularization procedure is applied to normalize the eigenvalues of the operators: we define $Z'_\rmB = \tilde Z^*_\rmB \abs{\tilde Z^*_\rmB}^{-1}$ where $\tilde Z^*_\rmB$ is $\tilde Z_\rmB$ with its zero eigenvalues changed to $1$.
The same procedure is applied to define $X'_\rmB$.
Alice's operators are unchanged because they are already unitary: $Z'_\rmA = \tilde Z_\rmA$ and $X'_\rmA = \tilde X_\rmA$.
As a side-note, we point out a slight mistake in \cite{mckague_robust_2012} concerning the above regularization procedure, see \apdxref{shortcomings}.

Second, we observe, as mentioned earlier and as pointed out in \cite{yang_robust_2013}, that the existence of a SOS decomposition for $\ibar_\alpha$ of the form \eqref{eq:sos} implies that any state $\ket{\tilde\psi}$ and operators $\tilde A_x$ and $\tilde B_y$ achieving the quantum bound $\imax_\alpha$ must obey the relations $\tilde P_i \ket{\tilde\psi} = 0$.
Moreover, this property is robust: if instead the expectation value is $\mean{\tilde \iop_\alpha} = \imax_\alpha-\epsilon$, then we have $\norm{\tilde P_i \ket{\tilde \psi}} \le \sqrt{\epsilon}$.
This observation applied to the two SOS \eqref{eq:sos1} and \eqref{eq:sos2} implies the following identities up to robust error terms:
\begin{subequations}
\label{eq:circuitids}
\begin{gather}
\label{eq:circuitids1}
(\tilde Z_\rmA - \tilde Z_\rmB) \ket{\tilde\psi} = 0 \eqp, \\
(\sin\theta \tilde X_\rmA (\id + \tilde Z_\rmB) - \cos\theta \tilde X_\rmB (\id - \tilde Z_\rmA)) \ket{\tilde\psi} = 0 \eqp, \\
(\tilde Z_\rmA \tilde X_\rmA - \tilde X_\rmA \tilde Z_\rmA) \ket{\tilde\psi} = 0 \eqp.
\end{gather}
\end{subequations}
For instance, identity \eqref{eq:circuitids1} in the case $\alpha=0$ is easily seen to follow from the term $P_1=(-Z_A+Z_B)$ present in the SOS decomposition \eqref{eq:chshsos2}.
The three identities \eqref{eq:circuitids} for arbitrary $\alpha$ follow in a similar, though more involved, manner from the two SOS \eqref{eq:sos1} and \eqref{eq:sos2}, see \apdxref{selftest}.

The third step of the proof then examines the action of the isometry on a state and observables satisfying the algebraic relations \eqref{eq:circuitids}.
It establishes that in this case the relations \eqref{eq:selftest} necessarily hold between the inputs and outputs of the isometry.
As in \cite{yang_robust_2013}, a lengthy series of triangle inequalities is needed to go from the SOS bounds on $\norm{P_i \ket{\tilde\psi}}$ to the robustness bounds for \eqref{eq:circuitids} and finally to \eqref{eq:selftest}.
We refer to \apdxref{selftest} for the derivation, where explicit bounds in $\bigo(\sqrt{\epsilon})$ for all the identities in \eqref{eq:selftest} are determined.

\section{Discussion}
In this work, we provided tools to simplify the search for sums-of-squares decompositions for Bell operators, exploiting their symmetries and the knowledge of systems that maximally violate the associated Bell inequalities.
We applied this approach to find two SOS decompositions for the family of tilted CHSH inequalities of \cite{acin_randomness_2012} (as well as some extra decompositions in the special case of the CHSH inequality).
We made use of these new SOS decompositions to complete and extend the proof of Yang and Navascués \cite{yang_robust_2013} by showing that a close-to-maximal quantum violation of a tilted CHSH inequality provides a robust self-test for the reference state and measurement operators associated with the inequality.

The general form $C \sqrt{\epsilon}$ of our distance bounds is optimal in the noise parameter $\epsilon$ in the sense that a larger exponent for $\epsilon$ would contradict the hypothesis $\mean{\ibar_\alpha} \le \epsilon$.
These distance bounds, though, become very sensitive to noise as the entanglement diminishes.
This is not surprising, as taking the entanglement parameter $\theta$ close to zero takes the extremal quantum behaviour for the $\iop_\alpha$ inequality closer to the local set.

The reference systems for the tilted CHSH inequalities are particularly relevant to randomness generation, as explained in \cite{acin_randomness_2012}.
Our self-testing statement may therefore be useful in establishing device-independent protocols using partially entangled states as a resource for randomness expansion.

We hope that a similar approach to finding SOS decompositions for Bell inequalities will find further applications in robust property testing in different systems, for instance with higher-dimensional reference Hilbert spaces.

\begin{acknowledgments}
We acknowledge financial support from the European Union under the project QALGO, from the F.R.S.-FNRS under the project DIQIP, and from the Brussels-Capital Region through a BB2B grant.
S. P. is a Research Associate of the Fonds de la Recherche Scientifique F.R.S.-FNRS (Belgium).
C. B. acknowledges funding from the F.R.S.-FNRS through a Research Fellowship.
\end{acknowledgments}

\input{appendices.tex}

\bibliography{ref.bib}

\end{document}

%% file: appendices.tex
\newcommand{\mref}{\relax} %does nothing, indicates a reference to the main text

\appendix

\section{Robust self-test}
\label{apdx:selftest}

\subsection{Introduction}
In this section, we show how our SOS results can be used to provide a robust self-test of the systems that maximally violate the $\iop_\alpha$ inequalities.
We work in the following setting: two isolated parties, Alice and Bob, hold black boxes that each take one of two inputs, respectively $x$ and $y$, in $\{0,1\}$, and each return one of two outputs, $a$ and $b$, in $\{+1, -1\}$.
They wish to make sure that their boxes share the partially entangled qubit pair $\ket{\psi} = \cos\theta \ket{00} + \sin\theta \ket{11}$ and that the observables characterizing their measurements on their share of the state are given in terms of the Pauli operators as
\begin{IEEEeqnarray}{-rL'rL}
\IEEEyesnumber\phantomsection\label{eq:a:maxops}
\IEEEyessubnumber
   A_0 &= \sigma_z
&  A_1 &= \sigma_x
\\
\IEEEyessubnumber
   B_0 &= \cos\mu \>\sigma_z + \sin\mu \>\sigma_x
&  B_1 &= \cos\mu \>\sigma_z - \sin\mu \>\sigma_x
\end{IEEEeqnarray}
with $\tan\mu = \sin 2\theta$.
In reality, the boxes hold the physical state $\ket{\tilde\psi}$ and the observables are the hermitian and dichotomic (i.e.\ of eigenvalues $\pm 1$) operators $\tilde A_x$ and $\tilde B_y$ -- this is general because there is no assumption made on the dimensionality of the Hilbert space, which can thus be extended to purify the state and make the measurements projective.

We follow the framework of McKague et al.\ \cite{mckague_robust_2012} and say that a Bell expression provides a robust self-test for a reference system if, for a violation of the corresponding Bell inequality that is $\epsilon$-close to the quantum maximum (we say that the system satisfies the self-testing criterion), the state and measurements that give rise to this violation are close to the reference, with the error vanishing as $\epsilon$ goes to zero.
Formally, this is stated as the existence of a local isometry with respect to Alice and Bob that takes $\ket{\tilde\psi}$ and the action of their observables $\tilde A_x$ and $\tilde B_y$ to a system close to the reference $\ket\psi$ and $A_x$, $B_y$, in tensor product with uncorrelated degrees of freedom.
The precise mathematical statement is the following: assuming that the self-testing criterion holds, there exists a local isometry $\Phi = \Phi_\rmA \otimes \Phi_\rmB$ and a state $\ket\junk$ such that
\begin{equation}
\norm{ \Phi(\tilde A_x \otimes \tilde B_y \ket{\tilde\psi}) - \ket{\junk} \otimes (A_x \otimes B_y) \ket{\psi} } \leq e_{xy}(\epsilon)
\label{eq:a:selftest}
\end{equation}
for all $x,y \in \{-1,0,1\}$ (where operators with subscript $-1$ refer to the identity), and $\lim_{\epsilon \to 0} e_{xy}(\epsilon) = 0$.
This is understood as meaning that there exists a procedure (that need not be accessible experimentally) that Alice and Bob can follow locally to perform a change of basis after which the reference state and operators (or something close) are found in a four-dimensional subspace of the global Hilbert space.

We show in this section a robust self-test for the ideal reference system highlighted above for $\iop_\alpha$ for any $\alpha \in \cointerval{0,2}$, with self-testing criterion $\bra{\tilde\psi} \iop_\alpha \ket{\tilde\psi} = \imax - \epsilon$ (with $\iop_\alpha$ expressed in terms of $\tilde A_x$ and $\tilde B_y$), leading to explicit self-testing bounds $e_{xy} \in O(\sqrt\epsilon)$.
This result uses the same techniques as \cite{yang_robust_2013} to achieve the self-test of the state (statement \eqref{eq:a:selftest} for $x=y=-1$) and extends the reasoning to the self-test of operators.

Following \cite{mckague_robust_2012,yang_robust_2013}, we define the following operators, essentially an inversion of \eqref{eq:a:maxops} for the physical operators:
\begin{IEEEeqnarray}{rL"rL}
\IEEEyesnumber\phantomsection\label{eq:a:xz}
\IEEEyessubnumber
   \tilde Z_\rmA &= \tilde A_0
&  \tilde X_\rmA &= \tilde  A_1
\label{eq:a:xza}
\eqp,\\
\IEEEyessubnumber
   \tilde Z_\rmB &= \frac{\tilde B_0 + \tilde B_1}{2\cos\mu}
&  \tilde X_\rmB &= \frac{\tilde B_0 - \tilde B_1}{2\sin\mu}
\label{eq:a:xzb}
\eqp.
\end{IEEEeqnarray}
We regularize these operators into unitaries to use them in the isometry.
The procedure is as follows: start from the physical observable, for example $\tilde Z_\rmB$, and change all zero eigenvalues to one, resulting in a new hermitian operator $Z^*_\rmB$.
Then, normalize the eigenvalues by defining $Z'_\rmB = \tilde Z^*_\rmB \abs{\tilde Z^*_\rmB}^{-1}$.
This last operator is by construction unitary, self-adjoint, commutes with $\tilde Z_\rmB$, and has the property that $Z'_\rmB \tilde Z_\rmB = \abs{\tilde Z_\rmB}$.
Similarly, we define $X'_\rmB$, $Z'_\rmA$ and $X'_\rmA$, noting that Alice's regularized $X'_\rmA$ and $Z'_\rmA$ operators actually coincide with the physical operators $\tilde X_\rmA$ and $\tilde X_\rmB$.

We now define the self-testing isometry by representing it in circuit form in Figure~\mref{\ref{fig:circuit}}.
In general, the action of the isometry on the physical state $\ket{\tilde\psi}$ is
\begin{equation}
\begin{split}
\Phi(\ket{\tilde\psi}) = \frac{1}{4}
\bigl[&
                (\id+Z'_\rmA) (\id+Z'_\rmB) \ket{\tilde\psi} \ket{00}
\\& +
X'_\rmA         (\id-Z'_\rmA) (\id+Z'_\rmB) \ket{\tilde\psi} \ket{10}
\\& +
        X'_\rmB (\id+Z'_\rmA) (\id-Z'_\rmB) \ket{\tilde\psi} \ket{01}
\\& +
X'_\rmA X'_\rmB (\id-Z'_\rmA) (\id-Z'_\rmB) \ket{\tilde\psi} \ket{11}
\bigr]
\eqp.
\end{split}
\label{eq:a:isomout}
\end{equation}

In this form, the motivation for this isometry is readily understood.
In the reference system with state $\ket\psi$ and operators \eqref{eq:a:maxops}, the $\tilde X$ and $\tilde Z$ operators above (and their regularizations) are the Pauli operators $\sigma_x$ and $\sigma_z$.
Then, $\Phi(\ket\psi)$ is easily seen to be $\ket{00}\ket\psi$, where the partially entangled qubit pair $\ket\psi$ has been extracted onto the ancilla qubits, leaving behind the state $\ket{00}$ in the physical register, unentangled with the ancilla.
Moreover, by manipulating \eqref{eq:a:isomout} using the operators' properties, it is easily seen in this ideal setting that acting on the input with the measurement operators is equivalent to acting on the corresponding ancillae with the reference operators.

Our goal is to extend this statement to all other systems that maximally violate the $\iop_\alpha$ inequality, and show its robustness to noise by deriving \eqref{eq:a:selftest}.
To do this, we need to show that the self-testing criterion implies some robust properties on the action of the physical operators on the state.

\subsection{Application of our SOS decompositions to robust self-testing}
We first start by reformulating our two SOS decompositions and show that, even though they contain only four independent terms, they still provide a way to test the entire five-dimensional space of relations that we defined as we identified candidates for SOS decompositions in Eqs.~\mref{\eqref{eq:soscandidates}}.
Defining the following polynomials in $A_x, B_y$:
\begin{subequations}
\label{eq:a:s}
\begin{align}
S_1 &= \ibar_\alpha
\\
S_2 &= \alpha A_1 - \mathcal S'
\\
S_3 &= 2 A_0 - \imax_\alpha \frac{B_0+B_1}{2} + \frac{\alpha}{2} \mathcal S''
\\
S_4 &= 2 A_1 - \imax_\alpha \frac{B_0-B_1}{2} + \frac{\alpha}{2} \mathcal S'''
\end{align}
\end{subequations}
with
\begin{align}
\mathcal S'   &= A_0 (B_0-B_1) + A_1 (B_0+B_1) \eqp, \\
\mathcal S''  &= A_0 (B_0+B_1) - A_1 (B_0-B_1) \eqp, \\
\mathcal S''' &= A_0 (B_0-B_1) - A_1 (B_0+B_1) \eqp,
\end{align}
the two SOS decompositions of $\ibar_\alpha$ are then
\begin{equation}
\label{eq:a:gensos}
\ibar_\alpha = \frac{1}{2\imax_\alpha} (S_1^2 + S_2^2) = \frac{1}{2\imax_\alpha} (S_3^2 + S_4^2) \eqp.
\end{equation}
We can also express the $S_i$ polynomials as $\vect s_i \cdot \vect V$ with $\vect V$ defined in \mref{\eqref{eq:vops}} and $\vect s_i$ expressed in terms of the basis vectors \mref{\eqref{eq:newr}}:
\begin{subequations}
\begin{align}
  \vect s_1 &= c \,\vect r_1 - \vect r_2 - \vect r_3
\\\vect s_2 &= -\vect r_5 \eqp,
\\\vect s_3 &= -\sqrt{1+s^2} \,\vect r_1 + \frac{c}{\sqrt{1+s^2}} (\vect r_2 - \vect r_3) \eqp,
\\\vect s_4 &= \frac{-2}{\sqrt{1+s^2}} \,\vect r_4 + \frac{c}{\sqrt{1+s^2}} \,\vect r_5 \eqp.
\end{align}
\end{subequations}

The four $S_i$ polynomials above do not span the entire candidate subspace $\vspan(\{R_i\})$ that we identified in Eqs.~\mref{\eqref{eq:soscandidates}}.
Indeed, while $s_2$ and $s_4$ generate the parity isotypical subspace spanned by $r_4$ and $r_5$, the other isotypical subspace is not spanned by the other two vectors.
For example, $r_3$ cannot be decomposed in terms of $s_1$ and $s_3$ alone.
We note however that left multiplication by $A_1$ takes an operator from the parity to the identity representation subspaces and vice versa.
In fact, the simple relation $R_3 = A_1 R_4$ holds.
Because $R_3$ is not a linear combination of the $S_i$ operators while $R_4$ is, we define $S_5 = R_3$ (and $\vect s_5 = \vect r_3$), which is then expressed in terms of the other four $S_i$ as
\begin{equation}
S_5 = A_1 \left(\frac{-c}{2} S_2 - \frac{\sqrt{1+s^2}}{2} S_4 \right) \eqp.
\label{eq:a:s5}
\end{equation}

Suppose now that the physical state $\ket{\tilde\psi}$ and observables $\tilde A_x, \tilde B_y$ satisfy the self-testing criterion with error $\epsilon$, i.e., the expectation value $\mean{\tilde \iop_\alpha}$ for the $\iop_\alpha$ Bell expression is such that $\mean{\tilde \iop_\alpha} = \imax_\alpha - \epsilon$.
Denote as $\iop_\alpha$ the Bell operator defined in terms of the observables $\tilde A_x$ and $\tilde B_y$, and let $\{P_i\}$ be a set of polynomials in these observables such that $\imax_\alpha \id - \iop_\alpha \equiv \ibar_\alpha = \sum_i P_i^\dag P_i^{\vphantom\dag}$.
Then, the self-testing criterion directly implies the bounds $\bra{\tilde\psi} P_i^\dag P_i^{\vphantom\dag} \ket{\tilde\psi} \le \epsilon$ or, equivalently, $\norm{P_i \ket{\tilde\psi}} \le \sqrt\epsilon$.
Hence, given a sum-of-squares decomposition of $\ibar_\alpha$, the action of the observables on the state is constrained for any state satisfying the self-testing criterion.

The SOS decompositions found in this article are valid for any $A_x$ and $B_y$ that are hermitian and dichotomic, where $A_x$ and $B_y$ commute.
By their definition and the assumption that Alice and Bob are separate, the physical observables $\tilde A_x$ and $\tilde B_y$ match these properties.
Hence, we can derive from our two SOS decompositions some useful bounds on the action of the observables on the state.

In our SOS decompositions \eqref{eq:a:gensos}, the $P_i$ operators are $(2 \imax_\alpha)^{-1/2} S_i$.
We thus let $\delta = \sqrt{2\imax_\alpha} \sqrt\epsilon$ so that the self-testing criterion implies $\norm{S_i \ket{\tilde\psi}} \le \delta$ for $i=1,2,3,4$.
From this, we deduce a similar bound that we will need for the action of $S_5$.
Because $\tilde A_1$ is unitary and appears as a left multiplier in $S_5$, we can drop it from the norm $\norm{S_5 \ket{\tilde\psi}}$ to find the norm of a linear combination of $S_2$ and $S_4$ acting on $\ket{\tilde\psi}$.
We then use the triangle inequality to bound this by $\bigo(\sqrt\epsilon)$:
\begin{align}
\norm{S_5 \ket{\tilde\psi}}
  &= \Norm{\frac{-c}{2} S_2 \ket{\tilde\psi} - \frac{\sqrt{1+s^2}}{2} S_4 \ket{\tilde\psi}}
\\&\le \frac{c}{2} \norm{S_2 \ket{\tilde\psi}} + \frac{\sqrt{1+s^2}}{2} \norm{S_4 \ket{\tilde\psi}}
\\&\le \frac{c+\sqrt{1+s^2}}{2} \delta \eqp.
\label{eq:a:s5bound}
\end{align}

Now that we have $O(\delta) = O(\sqrt\epsilon)$ bounds on $\norm{S_i \ket{\tilde\psi}}$ for all $i$, by using the triangle inequality we can robustly certify the action of the entire subspace of operators in $\mathcal S_{1+AB}$ that vanish in the ideal setting, which proves to be very useful in showing the effectiveness and robustness of the self-testing isometry.
Furthermore, we can extend this space beyond $\mathcal S_{1+AB}$ by left-multiplication by bounded operators as was done for $S_5$, which we will use in the self-test of measurements.

Recalling our general SOS decompositions, we notice that the four $S_i$ polynomials in addition to the $S_5$ polynomial (all formulated in terms of $\tilde A_x$ and $\tilde B_y$) can be linearly combined to form the following operators, used in \cite{yang_robust_2013} to prove the self-test:
\begin{IEEEeqnarray}{L}
\tilde Z_\rmA - \tilde Z_\rmB
=
\frac{-c}{2s^2} S_1 - \frac{\sqrt{1+s^2}}{2s^2} S_3 - \frac{c}{s^2} S_5
\eqp,\\
\sin\theta \tilde X_\rmA \left(\id + \tilde Z_\rmB\right) - \cos\theta \tilde X_\rmB \left(\id - \tilde Z_\rmA\right)
\IEEEnonumber\\\qquad =
\frac{\sqrt{1+s^2}}{8\sin\theta} \left( (c-2) S_2 + \sqrt{1+s^2} S_4 \right)
\eqp.
\label{eq:a:projoperator}
\end{IEEEeqnarray}
In anticipation to the measurement self-test, we add to this list the anticommutator between Alice's two observables.
This polynomial is not in $\mathcal S_{1+AB}$, but we can use left-multiplication to decompose it in terms of $S_i$:
\begin{IEEEeqnarray*}{-lL}
\IEEEeqnarraymulticol{2}{L}{
\tilde Z_\rmA \tilde X_\rmA + \tilde X_\rmA \tilde Z_\rmA =
}
\\
\qquad \frac{\sqrt{1+s^2}}{4s^2} \Bigl[
-2 S_2
&+ A_0 \left(-c S_2 + \sqrt{1+s^2} S_4\right)
\\
&+ A_1 \left( c S_1 + \sqrt{1+s^2} S_3\right)
\Bigr] \eqp.
\IEEEyesnumber
\end{IEEEeqnarray*}
We note that the anticommutator $\{ \tilde Z_\rmB, \tilde X_\rmB \}$ is zero from definition \eqref{eq:a:xzb}.
We will also need the following, which we write in terms of $S_5$ for the sake of briefness:
\begin{equation}
\tilde X_\rmB - \frac{1}{s} \tilde X_\rmA ( \id - c \tilde Z_\rmB ) = \frac{\sqrt{1+s^2}}{2s} A_1 S_5 \eqp.
\end{equation}

We now write down the explicit robustness bounds derived from the relations above in the same fashion as \eqref{eq:a:s5bound}:
\begin{gather}
\label{eq:a:delta1}
\norm{(\tilde Z_\rmA - \tilde Z_\rmB) \ket{\tilde\psi}}
\le \delta_1
\eqp, \\
\label{eq:a:delta2}
\Norm{\bigl(\sin\theta \tilde X_\rmA (\id + \tilde Z_\rmB) - \cos\theta \tilde X_\rmB (\id - \tilde Z_\rmA)\bigr) \ket{\tilde\psi}}
\le \delta_2
\eqp, \\
\label{eq:a:delta5}
\Norm{\bigl(\tilde X_\rmB - s^{-1} \tilde X_\rmA ( \id - c \tilde Z_\rmB )\bigr) \ket{\tilde\psi}}
\le \delta_5
\eqp, \\
\label{eq:a:deltaaa}
\norm{(\tilde Z_\rmA \tilde X_\rmA + \tilde X_\rmA \tilde Z_\rmA) \ket{\tilde\psi}} \le \delta_\rma^\rmA
\eqp,
\end{gather}
with
\begin{align}
\label{eq:a:delta1def}
\delta_1 &= (1+c)\frac{c+\sqrt{1+s^2}}{2s^2} \delta
\eqp, \\
\label{eq:a:delta2def}
\delta_2 &= \frac{\sqrt{1+s^2}}{8\sin\theta} \left((2-c) + \sqrt{1+s^2} \right) \delta
\eqp, \\
\label{eq:a:delta5def}
\delta_5 &= \frac{\sqrt{1+s^2} \left(c+\sqrt{1+s^2}\right)}{4s} \delta
\eqp, \\
\label{eq:a:deltaaadef}
\delta_\rma^\rmA &= \frac{\sqrt{1+s^2}}{2s^2} \left(1+c+\sqrt{1+s^2}\right) \delta
\eqp.
\end{align}

We now prove that the isometry defined earlier provides a robust self-test for our reference system.
This consists in applying a series of transformations on bounds \eqref{eq:a:delta1}--\eqref{eq:a:deltaaa} in order to reach bounds \eqref{eq:a:selftest}.
Concretely, in order to prove the self-testing bounds, the action of the unitary operators that constitute the isometry must be shown to be restricted by the self-testing criterion.
If there were no need for regularization, bounds \eqref{eq:a:delta1}--\eqref{eq:a:deltaaa} would directly apply to the isometry.
As we will show, the regularization procedure applied to define the unitaries in the isometry only introduces new error terms of the same order of $\bigo(\sqrt\epsilon)$ as the bounds for the unregularized operators.

\subsubsection{Self-testing the state}
We focus in this section on the claim of \cite{yang_robust_2013}, that is, the self-testing bound \eqref{eq:a:selftest} for the state (i.e., $x=y=-1$).

We start by showing an analogue of inequality \eqref{eq:a:delta1} for the regularized operators.
To do so, we use the triangle inequality to separate $\norm{(Z'_\rmB-Z'_\rmA)\ket{\tilde\psi}}$ into two terms.
The first one is bounded as follows:
\begin{align}
\norm{(Z'_\rmB - \tilde Z_\rmB) \ket{\tilde\psi}}
  &= \norm{Z'_\rmB (\id - \abs{\tilde Z_\rmB}) \ket{\tilde\psi}}
\\&= \norm{(\id - \abs{\tilde Z_\rmA \tilde Z_\rmB}) \ket{\tilde\psi}}
\\&\le \norm{(\id - \tilde Z_\rmA \tilde Z_\rmB) \ket{\tilde\psi}}
\\&\le \delta_1 \eqp.
\label{eq:a:zzb}
\end{align}
In the first equality, we used the identity $Z'_\rmB \abs{\tilde Z_\rmB} = \tilde Z_\rmB$ that we highlighted when defining the regularized operators.
In the second equality, we use the unitarity of $Z'_\rmB$ and the fact that the absolute value of an operator is unchanged by acting on the left with a unitary operator, here $\tilde Z_\rmA$.
The last inequality uses the unitarity of $\tilde Z_\rmA$ again to recover \eqref{eq:a:delta1}.
Recalling that $Z'_\rmA = \tilde Z_\rmA$, we thus have
\begin{align}
&\norm{(Z'_\rmB-Z'_\rmA) \ket{\tilde\psi}} \nonumber
\\
&\qquad \le \norm{(Z'_\rmB - \tilde Z_\rmB) \ket{\tilde\psi}} + \norm{(\tilde Z_\rmB - Z'_\rmA) \ket{\tilde\psi}}
\\
&\qquad \le 2 \delta_1 \eqp.
\label{eq:a:zzab}
\end{align}

We then prove a bound for $X'_\rmB$ similar to \eqref{eq:a:zzb}, which requires a different approach.
We note that
\begin{equation}
\cos^2\mu \, {\tilde Z_\rmB}^2 + \sin^2\mu \, {\tilde X_\rmB}^2 = \id \eqp,
\label{eq:a:tildeid}
\end{equation}
and use it in the following chain:
\begin{align}
&\norm{(X'_\rmB - \tilde X_\rmB) \ket{\tilde\psi}}
\nonumber
\\&\qquad = \norm{(\id-\abs{\tilde X_\rmB}) \ket{\tilde\psi}}
\\&\qquad \le \norm{(\id+\abs{\tilde X_\rmB})(\id-\abs{\tilde X_\rmB}) \ket{\tilde\psi}}
\\&\qquad = \cot^2(\mu) \norm{(\id-{\tilde Z_\rmB}^2) \ket{\tilde\psi}} \eqp.
\\&\qquad = \cot^2(\mu) \norm{(\id+{\tilde Z_\rmA \tilde Z_\rmB}) (\id-{\tilde Z_\rmA \tilde Z_\rmB}) \ket{\tilde\psi}}
\\&\qquad \le \cot^2(\mu) (1+(\cos\mu)^{-1}) \delta_1
\equiv \delta_4
\label{eq:a:xxb}
\end{align}
The first equality uses unitarity and the property $X'_\rmB \abs{\tilde X_\rmB} = \tilde X_\rmB$.
The first inequality uses the operator inequality $\id + \abs{\tilde X_\rmB} \succeq \id$.
The second equality uses \eqref{eq:a:tildeid}.
The last inequality uses \eqref{eq:a:tildeid} again to put a bound on $\norm{\tilde Z_\rmB}_\infty$.

We now turn to the self-testing statement.
In the isometry output \eqref{eq:a:isomout}, because $Z'_\rmA$ and $Z'_\rmB$ have near-identical action over the state by bound \eqref{eq:a:zzab}, the dichotomicity of the regularized operators makes the two middle terms approximately vanish.
In the first and last terms, for the same reason, the projectors $(\id\pm Z'_\rmA)/2$ and $(\id\pm Z'_\rmB)/2$ are nearly identical, and idempotence can be used.
In the first term, the error introduced by this approximation is bounded as follows:
\begin{align}
& \Norm{\left( \frac{\id+Z'_\rmA}{2}\frac{\id+Z'_\rmB}{2} - \frac{\id+Z'_\rmA}{2} \right) \ket{\tilde\psi}} \nonumber
\\
&\qquad =   \tfrac14 \Norm{ (\id+Z'_\rmA) \bigl( (\id+Z'_\rmB) - (\id+Z'_\rmA) \bigr) \ket{\tilde\psi}}
\\
&\qquad \le \tfrac12 \norm{Z'_\rmB - Z'_\rmA} \le \delta_1 \eqp.
\end{align}
We used the fact that $(\id+Z'_\rmA)^2 = 2 (\id+Z'_\rmA)$ in the first equality, and the operator bound $\norm{\id + Z'_\rmA}_\infty \le 2$ in the first inequality.
By the same reasoning, and using the fact that $X'_\rmA$ and $X'_\rmB$ are unitary to discard them from the norm, the fourth term leads to the same bound:
\begin{align}
\Norm{X'_\rmA X'_\rmB \left( \frac{\id-Z'_\rmA}{2}\frac{\id-Z'_\rmB}{2} - \frac{\id-Z'_\rmA}{2} \right) \ket{\tilde\psi}} \le \delta_1 \eqp.
\end{align}
The two middle terms are similarly bounded using the orthogonality of complementary projectors:
\begin{align}
& \Norm{\frac{\id \mp Z'_\rmA}{2}\frac{\id \pm Z'_\rmB}{2} \ket{\tilde\psi} - 0}
\nonumber
\\
&\qquad \le \tfrac14 \Norm{ (\id \mp Z'_\rmA) \bigl( (\id \pm Z'_\rmB) - (\id \pm Z'_\rmA) \bigr) \ket{\tilde\psi}}
\\
&\qquad \le \delta_1 \eqp.
\end{align}

Putting these bounds together, we deduce that replacing the isometry output with the state
\begin{equation}
\frac{\id+Z'_\rmA}{2} \ket{\tilde\psi} \ket{00} + X'_\rmA X'_\rmB \frac{\id-Z'_\rmA}{2} \ket{\tilde\psi} \ket{11}
\label{eq:a:isomout2}
\end{equation}
yields error terms bounded by $4 \delta_1$.
We note that this state is only approximately normalized.

We now take this approximation further to show that, as in the ideal case, the physical registers in the two terms are (approximately) proportional to the same state, which we call $\ket\junk$ and define as
\begin{equation}
\ket\junk = \beta^{-1} \frac{\id + Z'_\rmA}{2 \cos\theta} \ket{\tilde\psi} \eqp,
\label{eq:a:junk}
\end{equation}
where $\beta \ge 0$ is such that $\norm{\ket\junk} = 1$.
We will later show that $\beta \simeq 1$.
The first term in \eqref{eq:a:isomout2} is therefore $\beta \cos\theta \ket\junk \ket{00}$.
We show proportionality for the second term:
\begin{align}
&\Norm{X'_\rmA X'_\rmB \frac{\id-Z'_\rmA}{2} \ket{\tilde\psi} - \beta \sin\theta \ket\junk}
\nonumber
\\&\qquad =
\Norm{\left( X'_\rmA X'_\rmB \frac{\id-Z'_\rmA}{2} - \tan\theta \frac{\id+Z'_\rmA}{2} \right) \ket{\tilde\psi}}
\\&\qquad \le
\Norm{\left( X'_\rmB \frac{\id-Z'_\rmA}{2} - X'_\rmA \tan\theta \frac{\id+Z'_\rmB}{2} \right) \ket{\tilde\psi}}
\nonumber
\\&\qquad \quad +
\Norm{ \tan\theta \left( \frac{Z'_\rmB-Z'_\rmA}{2} \right) \ket{\tilde\psi}} \eqp.
\label{eq:a:isobound4}
\end{align}
The first equality uses the definition of $\ket\junk$.
The inequality uses the triangle inequality and the unitarity of $X'_\rmA$ on the first term.
The second term can be bounded by \eqref{eq:a:zzab}.
We can see that the first term is similar to bound \eqref{eq:a:delta2}, from which we now derive an equivalent for the regularized operators:
\begin{align}
&\Norm{\bigl( \cos\theta X'_\rmB (\id-Z'_\rmA) - \sin\theta X'_\rmA (\id+Z'_\rmB) \bigr) \ket{\tilde\psi}}
\nonumber
\\&\qquad \le
\Norm{\bigl( \cos\theta \tilde X_\rmB (\id-\tilde Z_\rmA) - \tilde X_\rmA \sin\theta (\id+\tilde Z_\rmB) \bigr) \ket{\tilde\psi}}
\nonumber
\\&\qquad \quad + \norm{ \cos\theta (\id-Z'_\rmA) (X'_\rmB - \tilde X_\rmB) \ket{\tilde\psi}}
\nonumber
\\&\qquad \quad + \norm{ \sin\theta X'_\rmA (\tilde Z_\rmB - Z'_\rmB) \ket{\tilde\psi}}
\\&\qquad \le
\delta_2 + 2 \delta_4 \cos\theta + \delta_1 \sin\theta
\equiv \delta'_2 \eqp.
\label{eq:a:delta2pdef}
\end{align}
The first inequality uses two triangle inequalities, the identities $Z'_\rmA = \tilde Z_\rmA$ and $X'_\rmA = \tilde X_\rmA$ and the commutation of these operators with Bob's.

All in all, \eqref{eq:a:isobound4} is thus bounded by $\tan\theta \delta_1 + \delta'_2 / \cos\theta$, and we end up with the following bound showing an approximate tensor product structure on the output state:
\begin{equation}
\Norm{\Phi(\ket{\tilde\psi}) - \beta \ket\junk \ket\psi} \le (4+\tan\theta) \delta_1 + \frac{\delta'_2}{2\cos\theta} \equiv \bar\delta \eqp.
\label{eq:a:deltabardef}
\end{equation}
The last error term to recover the state self-testing statement \eqref{eq:a:selftest} is to show that $\beta \simeq 1$.
To do so, we bound $1-\beta$ from above and below.
First, using the property that the isometry preserves the norm, we find
\begin{align}
1 &= \norm{\Phi(\ket{\tilde\psi})}
\\&\le \norm{\Phi(\ket{\tilde\psi}) - \beta \ket\junk \ket\psi} + \beta \norm{\ket\junk \ket\psi}
\\&\le \bar\delta + \beta \eqp.
\end{align}
Similarly, we have
\begin{align}
\beta &= \norm{\beta \ket\junk \ket\psi}
\\&\le \norm{\beta\ket\junk \ket\psi - \Phi(\ket{\tilde\psi})} + \norm{\Phi(\ket{\tilde\psi})}
\\&\le \bar\delta + 1 \eqp.
\end{align}
Put together, $\beta$ is constrained as $\abs{1-\beta} \le \bar\delta$.
Using this and \eqref{eq:a:deltabardef}, we finally complete the proof for the self-testing statement on the state
\begin{equation}
\Norm{\Phi(\ket{\tilde\psi}) - \ket\junk \ket\psi} \le 2 \bar\delta \eqp.
\label{eq:a:selfteststate}
\end{equation}

\subsubsection{Self-testing the measurements}
The proof for the self-testing statements involving operator actions in \eqref{eq:a:selftest} builds on the previous section.
As a first step, we give approximations on the output of the isometry when acting on the input with the \emph{regularized} operators, and then further approximate by substituting them with the nonregularized operators.

This is where we need the approximate anticommutation of $X'_\rmA$ and $Z'_\rmA$ when acting on the state, given by inequality \eqref{eq:a:deltaaa}.
We will also need to prove a similar anticommutation bound for $X'_\rmB$ and $Z'_\rmB$.
As we noted earlier, the nonregularized $\tilde X_\rmB$ and $\tilde Z_\rmB$ anticommute by definition, but the regularization procedure breaks this.
As a result, we can only prove that $\{\tilde X_\rmB, \tilde Z_\rmB\} \ket{\tilde\psi} \simeq 0$, which will require a number of successive approximations.

The use of anticommutation is made clear when looking at the action of the isometry \eqref{eq:a:isomout} on $X'_\rmA \ket{\tilde\psi}$ or $X'_\rmB \ket{\tilde\psi}$.
For example, (the reasoning for the other operator is the same):
\begin{align}
& \Phi(X'_\rmA \ket{\tilde\psi})
\nonumber
\\&\qquad =
\frac{1}{4} \bigl[
\begin{aligned}[t]
&               (\id+Z'_\rmB) (\id-X'_\rmA Z'_\rmA X'_\rmA) \ket{\tilde\psi} \ket{10}
\\& +
X'_\rmA         (\id+Z'_\rmB) (\id+X'_\rmA Z'_\rmA X'_\rmA) \ket{\tilde\psi} \ket{00}
\\& +
        X'_\rmB (\id-Z'_\rmB) (\id-X'_\rmA Z'_\rmA X'_\rmA) \ket{\tilde\psi} \ket{11}
\\& +
X'_\rmA X'_\rmB (\id-Z'_\rmB) (\id+X'_\rmA Z'_\rmA X'_\rmA) \ket{\tilde\psi} \ket{01}
\bigr]
\eqp.
\label{eq:a:isomxa}
\end{aligned}
\end{align}
The approximate anticommutation of $X'_\rmA$ and $Z'_\rmA$ means that $X'_\rmA Z'_\rmA X'_\rmA \ket{\tilde\psi} \simeq - Z'_\rmA \ket{\tilde\psi}$, and this output only differs from $\Phi(\ket{\tilde\psi})$ by the action of $\sigma_x$ on the first ancilla qubit, and a small error term.
On the other hand, it is easily seen that the isometry applied on $Z'_\rmA \ket{\tilde\psi}$ or $Z'_\rmB \ket{\tilde\psi}$ yields exactly the output $\Phi(\ket{\tilde\psi})$ with the $\sigma_z$ Pauli operator acting on the corresponding ancilla.

We now prove the anticommutation bound for Bob's operators.
Compared to the other bounds we have derived so far, this is not as immediate as it might seem;
indeed, while we know by \eqref{eq:a:zzb} and \eqref{eq:a:xxb} that the nonregularized and regularized $X$ and $Z$ operators on Bob's side are approximately interchangeable, this is only the case when they act on the physical state $\ket{\tilde\psi}$.
Thus, we can not deduce from this alone that $\{X'_\rmB,Z'_\rmB\} \ket{\tilde\psi} \simeq \{\tilde X_\rmB,\tilde Z_\rmB\} \ket{\tilde\psi} = 0$ because not all unitaries in the anticommutator act directly on the state.
Instead, what we do is to approximate the rightmost operator in each term of the anticommutator by its unnormalized counterpart by using \eqref{eq:a:zzb} and \eqref{eq:a:xxb}, and translate the action of $\tilde Z_\rmB$ and $\tilde X_\rmB$ into one that commutes with the leftmost operators using respectively \eqref{eq:a:delta1} and \eqref{eq:a:delta5}.
Then, the leftmost operators can in turn be approximated because they now act on the state directly.

The steps outlined above are carried out as follows:
\begin{align}
&\norm{\{X'_\rmB,Z'_\rmB\} \ket{\tilde\psi}}
\nonumber
%Z'_B -> Z_B -> Z_A and exact AB commutation
\\&\qquad\le
2 \delta_1 + \norm{(\tilde Z_\rmA + Z'_\rmB) X'_\rmB \ket{\tilde\psi}}
%Z_A + Z'_B <= 2 (unitaries), X'_B -> X_B then X_B -> other operator
\displaybreak[0]
\\&\qquad\le
2 \delta_1 + 2 \delta_4  + 2 \delta_5
\nonumber \\&\qquad\quad
+ s^{-1} \norm{(\tilde Z_\rmA + Z'_\rmB) \tilde X_\rmA (\id-c \tilde Z_\rmB) \ket{\tilde\psi}}
%AB commutation and commutation of Z'_B and Z_B
\displaybreak[0]
\\&\qquad=
2 \delta_1 + 2 \delta_4  + 2 \delta_5
\nonumber \\&\qquad\quad
+ s^{-1} \norm{(\id-c \tilde Z_\rmB) (\tilde Z_\rmA + Z'_\rmB) \tilde X_\rmA \ket{\tilde\psi}}
%bounding Z_B operator norm then A-B commutation
\displaybreak[0]
\\&\qquad\le
2 \delta_1 + 2 \delta_4  + 2 \delta_5
\nonumber \\&\qquad\quad
+ \frac{1+c/\!\cos\mu}{s} \norm{\tilde Z_\rmA \tilde X_\rmA + \tilde X_\rmA Z'_\rmB \ket{\tilde\psi}}
%Z'_B -> Z_B -> Z_A then A anticommutation
\displaybreak[0]
\\&\qquad\le
2 \delta_1 + 2 \delta_4  + 2 \delta_5 + \frac{1+c/\!\cos\mu}{s} (2 \delta_1 + \delta_\rma^\rmA)
\\&\qquad
\equiv \delta_\rma^\rmB
\eqp.
\end{align}
In the first inequality, we used \eqref{eq:a:zzab}.
The second inequality uses \eqref{eq:a:xxb} followed by \eqref{eq:a:delta5}, which lets us commute this approximation of $X'_\rmB$ to the left of the operator product in the first equality.
Next, in the third inequality, the operator bound $\norm{\tilde Z_\rmB}_\infty \le (\cos\mu)^{-1}$ that we used to derive \eqref{eq:a:xxb} is used again.
Finally, we use \eqref{eq:a:zzb} followed by Alice's anticommutation bound \eqref{eq:a:deltaaa} to reach the last inequality.

Thus, combining the anticommutation bounds with the regularization approximations for Bob's operators \eqref{eq:a:zzb} and \eqref{eq:a:xxb}, we find
\begin{align}
&\norm{\Phi(\tilde Z_\rmA \ket{\tilde\psi}) - \sigma_z^\rmA \Phi(\ket{\tilde \psi})} =   0 \eqp,
\label{eq:a:zabound}
 \\
&\norm{\Phi(\tilde X_\rmA \ket{\tilde\psi}) - \sigma_x^\rmA \Phi(\ket{\tilde \psi})} \le 2 \delta_\rma^{\rmA} \eqp,
\label{eq:a:xabound}
 \\
&\norm{\Phi(\tilde Z_\rmB \ket{\tilde\psi}) - \sigma_z^\rmB \Phi(\ket{\tilde \psi})} \le \delta_1 \eqp,
 \\
&\norm{\Phi(\tilde X_\rmB \ket{\tilde\psi}) - \sigma_x^\rmB \Phi(\ket{\tilde \psi})} \le \delta_4 + 2 \delta_\rma^{\rmB} \eqp.
\end{align}
In the last two bounds, we have also used the fact that $\Phi$ preserves the norms and is linear, such that for example
$\norm{\Phi(\tilde Z_\rmB \ket{\tilde\psi}) - \Phi(Z'_\rmB \ket{\tilde\psi})} = \norm{(\tilde Z_\rmB - Z'_\rmB)\ket{\tilde\psi}}$.

Our goal is to reach bounds from the joint action of the observables $\tilde A_x$ and $\tilde B_y$.
So far, we can only compute bounds for the action of one party at a time.
Indeed, to bound $\norm{\Phi(\tilde B_y \ket{\tilde\psi}) - B_y \Phi(\ket{\tilde\psi})}$ we can use definition \eqref{eq:a:maxops} for $B_y$, the triangle inequality, and the bounds above for the action of $\tilde Z_\rmB$ and $\tilde X_\rmB$.

We now show that with joint action of both parties, Alice's operator is easily dealt with.
First, consider $\Phi(\tilde A_0 \tilde B_y \ket{\tilde\psi})$.
As with \eqref{eq:a:zabound}, this is exactly the same as $A_0 \Phi(\tilde B_y \ket{\tilde\psi})$ because $\tilde A_0 = Z'_\rmA$ and $A_0 = \sigma_z^\rmA$.
Next is $\Phi(\tilde A_1 \tilde B_y \ket{\tilde\psi})$, for which it is easily seen that the reasoning that we used to reach \eqref{eq:a:xabound} is unchanged.
Indeed $A_1 = X'_\rmA$, and this state is identical to \eqref{eq:a:isomxa} with $\ket{\tilde\psi}$ replaced with $\tilde B_y \ket{\tilde\psi}$.
We use commutation between Alice and Bob to move $\tilde B_y$ to the left of $(\id \pm X'_\rmA Z'_\rmA X'_\rmA)$ and, because $\norm{\tilde B_y}_\infty = 1$, we find
\begin{equation}
\norm{\Phi(\tilde A_1 \tilde B_y \ket{\tilde\psi}) - A_1 \Phi(\tilde B_y \ket{\tilde\psi})} \le 2\delta_\rma^\rmA \eqp.
\end{equation}
We have thus showed that \eqref{eq:a:zabound} and \eqref{eq:a:xabound} are unchanged if we replace $\ket{\tilde\psi}$ with $\tilde B_y \ket{\tilde\psi}$ in both terms.

In addition to the above, we will use one last approximation to replace the isometry output $\Phi(\ket{\tilde\psi})$ with $\ket\junk \ket\psi$, at the cost of the additional error term of $2\bar\delta$ from \eqref{eq:a:selfteststate}.
We finally find the following bounds $e_{xy}(\epsilon)$ in the self-testing statement \eqref{eq:a:selftest}, where we define $e_{xy} = e'_{xy} + 2 \bar\delta$:
\begin{align}
e'_{-1,-1} &=
e'_{ 0,-1} = 0
\eqp,\\
e'_{ 1,-1} &= 2 \delta_\rma^\rmA
\eqp,\\
e'_{-1, 0} &=
e'_{-1, 1} =
e'_{ 0, 0} =
e'_{ 0, 1}   \nonumber\\
&= \delta_1 \cos\mu + (\delta_4 + 2 \delta_\rma^\rmB) \sin\mu
\eqp,\\
e'_{ 1, 0} &=
e'_{ 1, 1} =
2 \delta_\rma^\rmA + e'_{-1,0}
\eqp.
\end{align}

\section{Additional SOS for CHSH}
\label{apdx:extrachsh}
By guessing values for the parameters in \mref{\eqref{eq:chshsosmatrix}}, other nontrivial SOS decompositions than \mref{\eqref{eq:chshsos1}} and \mref{\eqref{eq:chshsos2}} can be found.

We report here a combination that leads to a SOS matrix of rank 4 which is extremal in the set represented in Figure~\mref{\ref{fig:circuit}}, i.e., it cannot be decomposed as a convex combination of SOS matrices of the same form as \mref{\eqref{eq:chshsosmatrix}}.
With the values $q=1/4$, $\mu=5/8$, $\lambda=1/4$, the square root of $M$ is rather well-behaved, and leads to the following SOS decomposition:
\begin{equation}
\ibar_0 = \frac{1}{8\sqrt{2}}
\begin{aligned}[t]
 \Bigl[
     &2 (Z_\rmA - Z_\rmB)^2 + 5 (X_\rmA - X_\rmB)^2
  \\ &+ 2 (Z_\rmA X_\rmB + X_\rmA Z_\rmB)^2
  \\ &+ (3 - 2 Z_\rmA Z_\rmB - X_\rmA X_\rmB)^2
 \Bigr] \eqp.
\end{aligned}
\end{equation}
An additional symmetry of the CHSH inequality can be exploited to reach an additional SOS decomposition: swapping $X$ and $Z$ (i.e., swapping the observables $A_0$ and $A_1$ and changing $B_1$ to $-B_1$) leaves $\ibar_0$ invariant, and therefore
\begin{equation}
\ibar_0 = \frac{1}{8\sqrt{2}}
\begin{aligned}[t]
 \Bigl[
     &2 (X_\rmA - X_\rmB)^2 + 5 (Z_\rmA - Z_\rmB)^2
  \\ &+ 2 (Z_\rmA X_\rmB + X_\rmA Z_\rmB)^2
  \\ &+ (3 - 2 X_\rmA X_\rmB - Z_\rmA Z_\rmB)^2
 \Bigr] \eqp,
\end{aligned}
\end{equation}
for which the parameter values are $q=1/4$, $\mu=1/4$, $\lambda=5/8$.

It is not clear whether either of these decompositions yields an easy generalization to the whole $\ibar_\alpha$ family.

\section{Shortcomings in previous results}
\label{apdx:shortcomings}

\subsection{CHSH self-test}
In their robust self-test proof for the maximal CHSH violation, McKague et al.\ \cite{mckague_robust_2012} introduce the isometry used in \cite{yang_robust_2013} and in the present article, with the same regularization construct we used on operators \eqref{eq:a:xz} to build the unitaries that make up the isometry.
The proof technique is the same as in this article: they identify from the self-testing criterion a series of constraints on the action of the observables in the system, that are then combined to form the self-testing statement.
However, in their self-test of the observables, the authors rely on the anticommutation relation $\{X'_\rmB, Z'_\rmB\} = 0$ on Bob's regularized operators.
As we noted, although $\tilde X_\rmB$ and $\tilde Z_\rmB$ (in our notation) anticommute by definition, this property is lost to regularization.
Indeed, the nonzero eigenvalues of anticommuting operators come in pairs of opposite sign, which means that nonsingular anticommuting operators do not exist in odd-dimensional Hilbert spaces.
In fact, if $B_0$ and $B_1$ do not each have as many $+1$ as $-1$ eigenvalues, their eigenspaces must share a nontrivial intersection, which will be an eigenspace for both $X'_\rmB$ and $Z'_\rmB$ (with nonzero eigenvalue because of regularization) where these two operators can therefore not anticommute.
Hence, in their proof in Appendix~B of \cite{mckague_robust_2012}, while $B_0' \pm B_1'$ anticommute, it is not true anymore when their zero eigenvalues are replaced with $1$.

However, this oversight only affects the final result by error terms of the same robustness order as they claim.
Indeed, as they show in the proof for their second self-testing criterion based on Mayers and Yao's work, the anticommutation of $X'_\rmB$ and $Z'_\rmB$ in front of the physical state $\ket{\psi'}$ can still be given a robustness bound.
This follows from the $\bigo(\sqrt\epsilon)$ bound on the anticommutation of Alice's $X'_\rmA$ and $Z'_\rmA$ operators and the $\bigo(\epsilon^{1/4})$ bounds on the replacement of Bob's $X'$ and $Z'$ operators by Alice's, which can be combined to transform Alice's anticommutation bound to one for Bob with $\bigo(\epsilon^{1/4})$ order.
Although this is worse than the $\epsilon_2 \in \bigo(\sqrt\epsilon)$ that they use in their Theorem~1, the final order in the self-testing bounds is unchanged by this correction because they already contain $\bigo(\epsilon^{1/4})$ terms from Alice's anticommutation bound.

\subsection{Partially entangled state self-test}
In their proof for the self-test statement \eqref{eq:a:selftest} on the state (i.e.\ $x = y = -1$), Yang and Navascu\'es introduce a SOS decomposition for $\ibar_\alpha$ different from ours.~\cite{yang_robust_2013}
They write the SOS polynomials $P_i$ as products $\vect q_i \cdot \vect V$ with $\vect V$ defined as in \mref{\eqref{eq:vops}}, and $\vect q_i$ decomposed in terms of five nine-dimensional vectors $\vect r_i$ different from the ones we defined in \mref{\eqref{eq:newr}}.
We reproduce here the $q_i$ they list in their article, with a change in $q_2$ which originally contained a typo that was communicated to us~\cite{yang_private}.
\begin{align}
\vect q_1 &= \frac{\gamma}{20\sqrt{2}} (\vect r_5 - \vect r_4) - \frac{2}{5} \vect r_1 \eqp, \\
\vect q_2 &= \frac{\sqrt{25 \sqrt{1+s^2} - 9 - \gamma^2/8}}{10s} (\vect r_1 + c \vect r_2 - c \vect r_3) \eqp, \\
\vect q_3 &= \frac{2\gamma - 25 c \sqrt{3-\overline{c}}}{30\sqrt{2}} \vect r_1 + \frac{3}{10} (\vect r_5 - \vect r_4) \eqp, \\
\vect q_4 &= \frac{35}{100} (\vect r_3 + \vect r_2) - \frac{5 c \sqrt{3-\overline{c}}}{14\sqrt{2}} \vect r_1 \eqp, \\
\vect q_5 &= \frac{\sqrt{49\gamma^2 + 9800 c \gamma \sqrt{3-\overline{c}} + \omega}}{420} \vect r_1 \eqp,
\label{eq:a:ynq5}
\end{align}
with
$\overline{c}=\cos(4\theta)$;
$\gamma=\sqrt{ (75+25\overline{c})\sqrt{6-2\overline{c}}-72}$; and
$\omega= 18125\cos(8\theta)-72500\cos(4\theta)-108706$,
Their SOS is then defined as
\begin{equation}
\ibar_\alpha = \sum_{i=1}^5 (\vect q_i \cdot \vect V)^\dag (\vect q_i \cdot \vect V) \eqp.
\label{eq:a:ynsos}
\end{equation}

This decomposition is problematic.
Indeed, the vector $\vect q_5$ is real only in the interval in $\theta$ (or equivalently in $\alpha$) where the expression inside the square root in \eqref{eq:a:ynq5} is positive.
This corresponds approximately to $\theta \in \ccinterval{0.07574, 0.73014}$, which is not the full interval $\ocinterval{0,\pi/4} = \ocinterval{0,0.78540}$.
Outside of that interval, a change of sign of the fifth term in \eqref{eq:a:ynsos} is actually required to recover the left-hand side.
However, this change means that the decomposition of $\ibar_\alpha$, while valid, is not a sum of squares anymore, and therefore it does not have the properties required for the self-test proof.
The value $\theta = \pi/8$ is within the validity interval; we represent the SOS in Figure~\mref{\ref{fig:pi8vol}} as a red dot.

Another issue with this decomposition is that the $\vect q_i$ vectors have a linear dependency that went unnoticed in the original article.
This can be seen in the fact that all five $\vect q_i$ only depend on $\vect r_4$ and $\vect r_5$ through their difference $\vect r_5 - \vect r_4$, which means that the $\vect q_i$ only span a four-dimensional subspace.
Hence, this is insufficient to certify the five operator identities in $\mathcal S_{1+AB}$.
Notably, forming the operator on the left hand side of \eqref{eq:a:projoperator} in the present article from their polynomials $\vect q_i \cdot \vect V$ requires a decomposition of $\vect r_5 + \vect r_4$ in terms of the $\vect q_i$ vectors as noted in the supplemental information to their article, which is not possible.

The linear dependency between the $\vect q_i$ is also visible on Figure~\mref{\ref{fig:pi8vol}}.
Indeed, the points on the boundary of this set correspond to singular SOS matrices $M$, as the nonsingular matrices, being strictly positive definite, do not saturate the inequalities that define the boundary.
Therefore, for this SOS decomposition, $M$ is singular (i.e.\ of rank at most $4$ here) and the $\vect q_i$ can therefore not be linearly independent.

\begin{table*}[ptb]
\newcommand{\CHSH}{\mathcal S}
\newcommand{\sbar}{\bar\CHSH}
\caption{\label{tab:a:chshvertexsos}
Vertex SOS decompositions for CHSH.}
\flushing% revtex4-1 command
\input{chshsos.tex}
\end{table*}

%% file: chshsos.tex
\paragraph*{\boldmath$\mathrm{C}_1$}
$\lambda = 0$,
%$\gamma = 1$,
$\mu = 0$,
$q = 0$
\begin{gather}
M = \frac{1}{2\sqrt{2}}
\begin{pmatrix}
0 & 0 & 0 & 0 & 0 \\
0 & 1 & 0 & 0 & 0 \\
0 & 0 & 1 & 0 & 0 \\
0 & 0 & 0 & 0 & 0 \\
0 & 0 & 0 & 0 & 0
\end{pmatrix}
\eqp, \\
\ibar_0 = \frac{1}{2\sqrt{2}} \left[ R_2^2 + R_3^2 \right]
      = \frac{1}{\sqrt{2}} \left[ (-\id + Z_\rmA Z_\rmB)^2 + (-\id + X_\rmA X_\rmB)^2 \right]
\eqp;
\end{gather}

\paragraph*{\boldmath$\mathrm{C}_2$}
$\lambda = 1$,
%$\gamma = 0$,
$\mu = 0$,
$q = 0$
\begin{gather}
M = \frac{1}{2\sqrt{2}}
\begin{pmatrix}
1 & 0 & 0 & 0 & 0 \\
0 & 0 & 0 & 0 & 0 \\
0 & 0 & 1 & 0 & 0 \\
0 & 0 & 0 & 0 & 0 \\
0 & 0 & 0 & 0 & 0
\end{pmatrix}
\eqp, \\
\ibar_0 = \frac{1}{2\sqrt{2}} \left[ R_1^2 + R_3^2 \right]
      = \frac{1}{\sqrt{2}} \left[ (-Z_\rmA + Z_\rmB)^2 + (-\id + X_\rmA X_\rmB)^2 \right]
\eqp;
\end{gather}

\paragraph*{\boldmath$\mathrm{C}_3$}

$\lambda = 0$,
%$\gamma = 1$,
$\mu = 1$,
$q = 0$
\begin{gather}
M = \frac{1}{2\sqrt{2}}
\begin{pmatrix}
0 & 0 & 0 & 0 & 0 \\
0 & 1 & 0 & 0 & 0 \\
0 & 0 & 0 & 0 & 0 \\
0 & 0 & 0 & 1 & 0 \\
0 & 0 & 0 & 0 & 0
\end{pmatrix}
\eqp, \\
\ibar_0 = \frac{1}{2\sqrt{2}} \left[ R_2^2 + R_4^2 \right]
      = \frac{1}{\sqrt{2}} \left[ (-\id + Z_\rmA Z_\rmB)^2 + (-X_\rmA + X_\rmB)^2 \right]
\eqp;
\end{gather}

\paragraph*{\boldmath$\mathrm{C}_4$}
$\lambda = 1$,
%$\gamma = 0$,
$\mu = 1$,
$q = 0$,
\begin{gather}
M = \frac{1}{2\sqrt{2}}
\begin{pmatrix}
1 & 0 & 0 & 0 & 0 \\
0 & 0 & 0 & 0 & 0 \\
0 & 0 & 0 & 0 & 0 \\
0 & 0 & 0 & 1 & 0 \\
0 & 0 & 0 & 0 & 0
\end{pmatrix}
\eqp, \\
\ibar_0 = \frac{1}{2\sqrt{2}} \left[ R_1^2 + R_4^2 \right]
      = \frac{1}{\sqrt{2}} \left[ (-Z_\rmA + Z_\rmB)^2 + (-X_\rmA + X_\rmB)^2 \right]
\eqp;
\end{gather}

\paragraph*{\boldmath$\mathrm{C}_{5}$}
$\lambda = 0$,
%$\gamma = 1/2$,
$\mu = 0$,
$q= 1/2$
\begin{gather}
M = \frac{1}{2\sqrt{2}}
\begin{pmatrix}
0 & 0   & 0   & 0 & 0   \\
0 & 1/2 & 1/2 & 0 & 0   \\
0 & 1/2 & 1/2 & 0 & 0   \\
0 & 0   & 0   & 0 & 0   \\
0 & 0   & 0   & 0 & 1/2
\end{pmatrix}
\eqp, \\
\ibar_0 = \frac{1}{4\sqrt{2}} \left[ \left(R_2 + R_3\right)^2 + R_5^2 \right]
      = \frac{1}{4\sqrt{2}} \left[ (-2\sqrt{2} \>\id + \ibar_0)^2 + 2 ( Z_\rmA X_\rmB + X_\rmA Z_\rmB )^2 \right]
      = \frac{1}{4\sqrt{2}} \left[ \ibar_0^2 + {\CHSH'}^2 \right]
\eqp,
\end{gather}
where $\CHSH' = A_0 (B_0-B_1) + A_1 (B_0+B_1)$.

%% file: sospaper.bbl
\begin{thebibliography}{16}%
\makeatletter
\providecommand \@ifxundefined [1]{%
 \@ifx{#1\undefined}
}%
\providecommand \@ifnum [1]{%
 \ifnum #1\expandafter \@firstoftwo
 \else \expandafter \@secondoftwo
 \fi
}%
\providecommand \@ifx [1]{%
 \ifx #1\expandafter \@firstoftwo
 \else \expandafter \@secondoftwo
 \fi
}%
\providecommand \natexlab [1]{#1}%
\providecommand \enquote  [1]{``#1''}%
\providecommand \bibnamefont  [1]{#1}%
\providecommand \bibfnamefont [1]{#1}%
\providecommand \citenamefont [1]{#1}%
\providecommand \href@noop [0]{\@secondoftwo}%
\providecommand \href [0]{\begingroup \@sanitize@url \@href}%
\providecommand \@href[1]{\@@startlink{#1}\@@href}%
\providecommand \@@href[1]{\endgroup#1\@@endlink}%
\providecommand \@sanitize@url [0]{\catcode `\\12\catcode `\$12\catcode
  `\&12\catcode `\#12\catcode `\^12\catcode `\_12\catcode `\%12\relax}%
\providecommand \@@startlink[1]{}%
\providecommand \@@endlink[0]{}%
\providecommand \url  [0]{\begingroup\@sanitize@url \@url }%
\providecommand \@url [1]{\endgroup\@href {#1}{\urlprefix }}%
\providecommand \urlprefix  [0]{URL }%
\providecommand \Eprint [0]{\href }%
\providecommand \doibase [0]{http://dx.doi.org/}%
\providecommand \selectlanguage [0]{\@gobble}%
\providecommand \bibinfo  [0]{\@secondoftwo}%
\providecommand \bibfield  [0]{\@secondoftwo}%
\providecommand \translation [1]{[#1]}%
\providecommand \BibitemOpen [0]{}%
\providecommand \bibitemStop [0]{}%
\providecommand \bibitemNoStop [0]{.\EOS\space}%
\providecommand \EOS [0]{\spacefactor3000\relax}%
\providecommand \BibitemShut  [1]{\csname bibitem#1\endcsname}%
\let\auto@bib@innerbib\@empty
%</preamble>
\bibitem [{\citenamefont {Brunner}\ \emph {et~al.}(2014)\citenamefont
  {Brunner}, \citenamefont {Cavalcanti}, \citenamefont {Pironio}, \citenamefont
  {Scarani},\ and\ \citenamefont {Wehner}}]{rmp}%
  \BibitemOpen
  \bibfield  {author} {\bibinfo {author} {\bibfnamefont {N.}~\bibnamefont
  {Brunner}}, \bibinfo {author} {\bibfnamefont {D.}~\bibnamefont {Cavalcanti}},
  \bibinfo {author} {\bibfnamefont {S.}~\bibnamefont {Pironio}}, \bibinfo
  {author} {\bibfnamefont {V.}~\bibnamefont {Scarani}}, \ and\ \bibinfo
  {author} {\bibfnamefont {S.}~\bibnamefont {Wehner}},\ }\href {\doibase
  10.1103/RevModPhys.86.419} {\bibfield  {journal} {\bibinfo  {journal} {Rev.
  Mod. Phys.}\ }\textbf {\bibinfo {volume} {86}},\ \bibinfo {pages} {419}
  (\bibinfo {year} {2014})}\BibitemShut {NoStop}%
\bibitem [{Note1()}]{Note1}%
  \BibitemOpen
  \bibinfo {note} {We assume implicitly that Alice's observables are of the
  form $A_x\otimes {\protect \mathbb I}$ and those of Bob of the form
  ${\protect \mathbb I}\otimes B_y$.}\BibitemShut {Stop}%
\bibitem [{\citenamefont {Ac\'in}\ \emph {et~al.}(2012)\citenamefont {Ac\'in},
  \citenamefont {Massar},\ and\ \citenamefont
  {Pironio}}]{acin_randomness_2012}%
  \BibitemOpen
  \bibfield  {author} {\bibinfo {author} {\bibfnamefont {A.}~\bibnamefont
  {Ac\'in}}, \bibinfo {author} {\bibfnamefont {S.}~\bibnamefont {Massar}}, \
  and\ \bibinfo {author} {\bibfnamefont {S.}~\bibnamefont {Pironio}},\ }\href
  {\doibase 10.1103/PhysRevLett.108.100402} {\bibfield  {journal} {\bibinfo
  {journal} {Phys. Rev. Lett.}\ }\textbf {\bibinfo {volume} {108}},\ \bibinfo
  {pages} {100402} (\bibinfo {year} {2012})}\BibitemShut {NoStop}%
\bibitem [{\citenamefont {Doherty}\ \emph {et~al.}(2008)\citenamefont
  {Doherty}, \citenamefont {Liang}, \citenamefont {Toner},\ and\ \citenamefont
  {Wehner}}]{doherty}%
  \BibitemOpen
  \bibfield  {author} {\bibinfo {author} {\bibfnamefont {A.~C.}\ \bibnamefont
  {Doherty}}, \bibinfo {author} {\bibfnamefont {Y.-C.}\ \bibnamefont {Liang}},
  \bibinfo {author} {\bibfnamefont {B.}~\bibnamefont {Toner}}, \ and\ \bibinfo
  {author} {\bibfnamefont {S.}~\bibnamefont {Wehner}},\ }in\ \href@noop {}
  {\emph {\bibinfo {booktitle} {Computational Complexity, 2008. CCC'08. 23rd
  Annual IEEE Conference on}}}\ (\bibinfo {organization} {IEEE},\ \bibinfo
  {year} {2008})\ pp.\ \bibinfo {pages} {199--210}\BibitemShut {NoStop}%
\bibitem [{\citenamefont {Navascu\'es}\ \emph {et~al.}(2007)\citenamefont
  {Navascu\'es}, \citenamefont {Pironio},\ and\ \citenamefont
  {Ac\'{\i}n}}]{npa}%
  \BibitemOpen
  \bibfield  {author} {\bibinfo {author} {\bibfnamefont {M.}~\bibnamefont
  {Navascu\'es}}, \bibinfo {author} {\bibfnamefont {S.}~\bibnamefont
  {Pironio}}, \ and\ \bibinfo {author} {\bibfnamefont {A.}~\bibnamefont
  {Ac\'{\i}n}},\ }\href {\doibase 10.1103/PhysRevLett.98.010401} {\bibfield
  {journal} {\bibinfo  {journal} {Phys. Rev. Lett.}\ }\textbf {\bibinfo
  {volume} {98}},\ \bibinfo {pages} {010401} (\bibinfo {year}
  {2007})}\BibitemShut {NoStop}%
\bibitem [{\citenamefont {Navascu\'es}\ \emph {et~al.}(2008)\citenamefont
  {Navascu\'es}, \citenamefont {Pironio},\ and\ \citenamefont
  {Ac\'in}}]{navascues_convergent_2008}%
  \BibitemOpen
  \bibfield  {author} {\bibinfo {author} {\bibfnamefont {M.}~\bibnamefont
  {Navascu\'es}}, \bibinfo {author} {\bibfnamefont {S.}~\bibnamefont
  {Pironio}}, \ and\ \bibinfo {author} {\bibfnamefont {A.}~\bibnamefont
  {Ac\'in}},\ }\href {\doibase 10.1088/1367-2630/10/7/073013} {\bibfield
  {journal} {\bibinfo  {journal} {New J. Phys.}\ }\textbf {\bibinfo {volume}
  {10}},\ \bibinfo {pages} {073013} (\bibinfo {year} {2008})}\BibitemShut
  {NoStop}%
\bibitem [{\citenamefont {Yang}\ and\ \citenamefont
  {Navascu\'es}(2013)}]{yang_robust_2013}%
  \BibitemOpen
  \bibfield  {author} {\bibinfo {author} {\bibfnamefont {T.~H.}\ \bibnamefont
  {Yang}}\ and\ \bibinfo {author} {\bibfnamefont {M.}~\bibnamefont
  {Navascu\'es}},\ }\href {\doibase 10.1103/PhysRevA.87.050102} {\bibfield
  {journal} {\bibinfo  {journal} {Phys. Rev. A}\ }\textbf {\bibinfo {volume}
  {87}},\ \bibinfo {pages} {050102} (\bibinfo {year} {2013})}\BibitemShut
  {NoStop}%
\bibitem [{\citenamefont {Mayers}\ and\ \citenamefont {Yao}(2004)}]{mayers}%
  \BibitemOpen
  \bibfield  {author} {\bibinfo {author} {\bibfnamefont {D.}~\bibnamefont
  {Mayers}}\ and\ \bibinfo {author} {\bibfnamefont {A.}~\bibnamefont {Yao}},\
  }\href@noop {} {\bibfield  {journal} {\bibinfo  {journal} {Quantum
  Information \& Computation}\ }\textbf {\bibinfo {volume} {4}},\ \bibinfo
  {pages} {273} (\bibinfo {year} {2004})}\BibitemShut {NoStop}%
\bibitem [{\citenamefont {Reichardt}\ \emph {et~al.}(2013)\citenamefont
  {Reichardt}, \citenamefont {Unger},\ and\ \citenamefont {Vazirani}}]{ruv}%
  \BibitemOpen
  \bibfield  {author} {\bibinfo {author} {\bibfnamefont {B.~W.}\ \bibnamefont
  {Reichardt}}, \bibinfo {author} {\bibfnamefont {F.}~\bibnamefont {Unger}}, \
  and\ \bibinfo {author} {\bibfnamefont {U.}~\bibnamefont {Vazirani}},\ }\href
  {\doibase 10.1038/nature12035} {\bibfield  {journal} {\bibinfo  {journal}
  {Nature}\ }\textbf {\bibinfo {volume} {496}},\ \bibinfo {pages} {456}
  (\bibinfo {year} {2013})}\BibitemShut {NoStop}%
\bibitem [{\citenamefont {Miller}\ and\ \citenamefont
  {Shi}(2014)}]{miller_robust_2014}%
  \BibitemOpen
  \bibfield  {author} {\bibinfo {author} {\bibfnamefont {C.~A.}\ \bibnamefont
  {Miller}}\ and\ \bibinfo {author} {\bibfnamefont {Y.}~\bibnamefont {Shi}},\
  }\href {http://arxiv.org/abs/1402.0489} {\bibfield  {journal} {\bibinfo
  {journal} {arXiv:1402.0489 [quant-ph]}\ } (\bibinfo {year}
  {2014})}\BibitemShut {NoStop}%
\bibitem [{\citenamefont {Almeida}\ \emph {et~al.}(2010)\citenamefont
  {Almeida}, \citenamefont {Bancal}, \citenamefont {Brunner}, \citenamefont
  {Ac\'{\i}n}, \citenamefont {Gisin},\ and\ \citenamefont {Pironio}}]{GYNI}%
  \BibitemOpen
  \bibfield  {author} {\bibinfo {author} {\bibfnamefont {M.~L.}\ \bibnamefont
  {Almeida}}, \bibinfo {author} {\bibfnamefont {J.-D.}\ \bibnamefont {Bancal}},
  \bibinfo {author} {\bibfnamefont {N.}~\bibnamefont {Brunner}}, \bibinfo
  {author} {\bibfnamefont {A.}~\bibnamefont {Ac\'{\i}n}}, \bibinfo {author}
  {\bibfnamefont {N.}~\bibnamefont {Gisin}}, \ and\ \bibinfo {author}
  {\bibfnamefont {S.}~\bibnamefont {Pironio}},\ }\href@noop {} {\bibfield
  {journal} {\bibinfo  {journal} {Physical review letters}\ }\textbf {\bibinfo
  {volume} {104}},\ \bibinfo {pages} {230404} (\bibinfo {year}
  {2010})}\BibitemShut {NoStop}%
\bibitem [{\citenamefont {{McKague}}\ \emph {et~al.}(2012)\citenamefont
  {{McKague}}, \citenamefont {Yang},\ and\ \citenamefont
  {Scarani}}]{mckague_robust_2012}%
  \BibitemOpen
  \bibfield  {author} {\bibinfo {author} {\bibfnamefont {M.}~\bibnamefont
  {{McKague}}}, \bibinfo {author} {\bibfnamefont {T.~H.}\ \bibnamefont {Yang}},
  \ and\ \bibinfo {author} {\bibfnamefont {V.}~\bibnamefont {Scarani}},\ }\href
  {\doibase 10.1088/1751-8113/45/45/455304} {\bibfield  {journal} {\bibinfo
  {journal} {J. Phys. A: Math. Theor.}\ }\textbf {\bibinfo {volume} {45}},\
  \bibinfo {pages} {455304} (\bibinfo {year} {2012})}\BibitemShut {NoStop}%
\bibitem [{\citenamefont {Sturm}(1999)}]{sedumi}%
  \BibitemOpen
  \bibfield  {author} {\bibinfo {author} {\bibfnamefont {J.}~\bibnamefont
  {Sturm}},\ }\href@noop {} {\bibfield  {journal} {\bibinfo  {journal}
  {Optimization Methods and Software}\ }\textbf {\bibinfo {volume} {11--12}},\
  \bibinfo {pages} {625} (\bibinfo {year} {1999})},\ \bibinfo {note} {version
  1.05 available from {\texttt{http://fewcal.kub.nl/sturm}}}\BibitemShut
  {NoStop}%
\bibitem [{\citenamefont {Gatermann}\ and\ \citenamefont
  {Parrilo}(2004)}]{gatermann_symmetry_2004}%
  \BibitemOpen
  \bibfield  {author} {\bibinfo {author} {\bibfnamefont {K.}~\bibnamefont
  {Gatermann}}\ and\ \bibinfo {author} {\bibfnamefont {P.~A.}\ \bibnamefont
  {Parrilo}},\ }\href {\doibase 10.1016/j.jpaa.2003.12.011} {\bibfield
  {journal} {\bibinfo  {journal} {J. Pure Appl. Algebra}\ }\textbf {\bibinfo
  {volume} {192}},\ \bibinfo {pages} {95} (\bibinfo {year} {2004})}\BibitemShut
  {NoStop}%
\bibitem [{\citenamefont {Bhatia}(2009)}]{bhatia_positive_2009}%
  \BibitemOpen
  \bibfield  {author} {\bibinfo {author} {\bibfnamefont {R.}~\bibnamefont
  {Bhatia}},\ }\href@noop {} {\emph {\bibinfo {title} {Positive Definite
  Matrices}}}\ (\bibinfo  {publisher} {Princeton University Press},\ \bibinfo
  {year} {2009})\BibitemShut {NoStop}%
\bibitem [{\citenamefont {Yang}(2014)}]{yang_private}%
  \BibitemOpen
  \bibfield  {author} {\bibinfo {author} {\bibfnamefont {T.~H.}\ \bibnamefont
  {Yang}},\ }\href@noop {} {}\bibinfo {howpublished} {personal communication}
  (\bibinfo {year} {2014})\BibitemShut {NoStop}%
\end{thebibliography}%
